%% file: main.tex
\documentclass[lettersize,journal]{IEEEtran}
\usepackage{amsmath,amsfonts,amssymb}
\usepackage{algorithm}
\usepackage{array}
\usepackage[caption=false,font=normalsize,labelfont=sf,textfont=sf]{subfig}
\usepackage{textcomp}
\usepackage{stfloats}
\usepackage{url}
\usepackage{verbatim}
\usepackage{graphicx}
\usepackage{cite}
\input{preamble.tex}
\usepackage{times,float}
\usepackage{xspace,syntonly,algpseudocode}
\usepackage{verbatim,multirow}
\usepackage[T1]{fontenc}
\usepackage[normalem]{ulem}

\newcommand{\revise}[1]{\textcolor{black}{#1}}

\newcommand{\shaohui}[1]{\textcolor{teal}{{SL: #1}}}

\begin{document}

\title{Watts vs. Bytes: Turning Data Centers into Grid Assets via Storage–Compute Co-Optimization}
\author{Shaohui Liu,~\IEEEmembership{Member,~IEEE}, Sungho Shin,~\IEEEmembership{Member,~IEEE}, Deepjyoti Deka,~\IEEEmembership{Senior Member,~IEEE}
\thanks{This research is funded by MIT Energy Initiative Future Energy System Center. All authors are with MIT. email: \{shaohuil,sushin,deepj87\}@mit.edu.}
}



\maketitle

\begin{abstract}
Data center interconnections increasingly face tighter peak-demand and ramp-rate limits while being expected to support grid operations. Satisfying these requirements calls for coordinated computing and energy controls, yet their joint operational and economic implications remain poorly understood. To tackle this problem, we formulate {a robust day-ahead co-optimization of} computing load scheduling, server dynamic voltage and frequency scaling (DVFS), and co-located battery energy storage system (BESS) dispatch. {The resulting mixed-integer linear program} hedges against uncertainty in fixed load and ancillary service deployment while enforcing interconnection limits on peak demand and ramp rate, ancillary service capacity commitments in reserve and flexible ramping{,} and workload execution constraints. Case studies using CAISO and PJM market data of a 100~MW data center with a 36~MWh/12~MW BESS show that workload scheduling, DVFS, and storage provide complementary flexibility. Under binding peak-load limits, increasing the schedulable workload share reduces mean daily operating cost by up to 20.7\%, and the daily value of storage more than doubles relative to operation under less restrictive limits. Under normal conditions, optimal BESS sizing is driven more by capital cost and cycling allowance than by energy duration alone. An 8~MW aggregate ancillary service commitment increases operational cost by only 0.4\%, whereas {reserve-only requirements become infeasible at commitments as small as 4~MW}. These findings show that coordinated computing and storage controls can support grid services economically under binding interconnection constraints while protecting workload delivery.
\end{abstract}


\section{Introduction}
Rapid growth in Generative AI is accelerating data center electricity demand. Accommodating this demand requires not only energy supply but also operating reserves, ramping capability, and regulation services. Because planning and procuring these resources require long lead times, several system operators have reported substantial interconnection delays for new data center loads~\cite{ercot2025plan,norris2025rethinking}. 

These pressures are prompting a broad discussion of new requirements on how large loads interconnect with the grid. One direction imposes operator-enforced limits on a data center's peak demand and ramping~\cite{brancucci2025flexible}; another proposes participation in peak shaving, demand response, and other price-responsive programs to reduce the need for reliability asset procurement, although recent North American Electric Reliability Corporation (NERC) guidance highlights uncertainty {regarding} their implementation~\cite{nerc2026gaps}. This motivates the central question of this paper:

\begin{quote}
 \emph{\textbf{In what form} should a data center offer flexibility to grid operations, \textbf{what internal resources} should it coordinate to deliver that flexibility, and \textbf{at what cost} to its own operations and quality of service?}
\end{quote}

These three dimensions define coupled modeling requirements. First, flexibility must be expressed through grid-operational quantities{,} such as reserve, flexible ramping product (FRP) or regulation commitments under peak power and ramp-rate limits at the point of data center interconnection~\cite{li2015flexible,ghaljehei2021day}. Second, deliverability requires coordination among workload scheduling, battery energy storage system (BESS) dispatch, and {data center} controls such as dynamic voltage and frequency scaling (DVFS), whose operational constraints jointly determine the feasible response~\cite{stewart2024bess,piga2024expanding}. Third, the economic impact of flexibility must account for worst-case energy procurement and ancillary service revenues as well as penalties and costs for workload-quality and BESS degradation~\cite{wang2018adjustable,cole2025cost,lauinger2025value}{.} We quantify these impacts by the change in optimized net operating cost relative to the corresponding no-service benchmark.

\paragraph*{Related work}
Classical data center power management emphasizes internal efficiency, provisioning, and server-level control~\cite{fan2007power,guenter2011managing}, whereas recent work examines grid-interactive operation~\cite{hasegawa2023demandresponse,googletrace2020} and demand flexibility as a means of facilitating faster interconnection~\cite{brancucci2025flexible}. Geographically coordinated load shifting and carbon-aware workload management are studied in~\cite{dvorkin2024agent,lin2023adapting,radovanovic2022carbon,zhang2020flexibility}, but these works emphasize network-wide or multi-site coordination rather than a single interconnection-constrained facility.

On the storage side, coordinated battery participation across electricity-market products is studied in~\cite{engels2020integration}; prior studies also examine the joint use of data center load and BESS for frequency regulation~\cite{guruprasad2017coupling} and the broader provision of ancillary services by data centers~\cite{wierman2014opportunities,fu2020assessments}. However, existing works {focus mainly on individual aspects rather than system-level coupling, and thus fail to reveal the value of} co-located BESS in multiple grid services while enforcing interconnection limits and coordinating flexible computing loads.

The robust virtual capacity curve (VCC) formulation for day-ahead planning in~\cite{hall2025carbon} is closest in spirit to our setting. We extend this perspective by jointly modeling co-located BESS, discrete DVFS, interconnection limits, and minimum market-participation requirements while explicitly accounting for workload-quality effects. More broadly, existing power-system flexibility models provide useful high-level abstractions~\cite{zhao2016flex,Riaz2022flex} for operations, but they do not directly capture the coupling between data center control choices and uncertain ancillary service deployment.

\paragraph*{Our approach}
We formulate a {robust} day-ahead co-optimization of workload scheduling, discrete DVFS, and co-located BESS dispatch for a data center subject to interconnection limits, and minimum ancillary service participation. Specifically, our formulation guarantees ancillary service delivery by the BESS and compliance with grid-operator-specified interconnection requirements for the {data center}. We consider reserve and flexible ramping product (FRP) under grid ancillary services because their hourly day-ahead timescale is consistent with the scheduling horizon; faster regulation is delegated to real-time controls and outside our purview. We model constraints and costs for each {data center} control (schedulable workloads, DVFS, and BESS) and use a robust formulation to hedge against forecast errors in fixed/non-schedulable loads and ancillary service deployment that are not exactly known at the day-ahead scheduling stage. We evaluate operating cost and computing service performance for our approach and further {validate} performance under unseen realizations.

\paragraph*{Contributions}
The main contributions of this paper are as follows:
\begin{itemize}[leftmargin=*]

\item We develop a flexibility-centric robust day-ahead co-optimization that coordinates the grid-facing balancing capability of BESS with complementary workload-scheduling and DVFS flexibility of {the data center}. The model guarantees delivery of scheduled reserve and FRP capacity by the BESS and enforces peak-demand and ramp-rate limits for every scenario trajectory, while accommodating aggregate or product-specific minimum-participation requirements. This integrated robust MILP is the principal methodological contribution.

\item Using real market and load data, we quantify how workload scheduling, BESS, and DVFS contribute to {data center} flexibility. In normal tested configurations, the 36~MWh/12~MW BESS and the nine-level DVFS set reduce mean operating cost by 2.0\% and 4.4\%, respectively, relative to their corresponding baselines. Under a binding 90~MW peak-load limit, increasing the schedulable-workload share from 5\% to 40\% reduces mean daily operating cost by 20.7\%, from \$302.0k/day to \$239.4k/day. Additionally, an obligatory 8~MW aggregate ancillary service commitment increases mean operating cost by only 0.4\% without reducing workload completion, while product-specific reserve obligations {become infeasible at commitments as small as 4~MW}. We further evaluate commercial BESS sizing under alternative cycling and capital-recovery assumptions.
\end{itemize}

\paragraph*{Organization}
The rest of this paper is organized as follows. \Cref{sec:problem_formulation} presents the joint data center and BESS formulation. \Cref{sec:case_studies} reports the representative dispatch, battery valuation and sizing, interconnection and discrete-DVFS sensitivities, and minimum market-participation study. \Cref{sec:conclusion} concludes the paper. Supplementary formulations, case-study details, ablation results, and nomenclature are provided in {\crefrange{app:explicit-formulation}{app:additional-ablations}}.


\section{Problem formulation}
\label{sec:problem_formulation}
We consider a single data center that participates in the day-ahead reserve and flexible ramping product (FRP) markets while respecting utility-imposed interconnection limits on peak load and ramping. The operator must commit a feasible schedule, including workload assignments, DVFS settings, BESS dispatch, and market capacity offers. We first state the notation, then present the overall robust co-optimization structure, followed by detailed models of each subsystem.

Throughout, we index time by $\mathcal{T}=\{1,\dots,T\}$ with $T=24$ hourly periods and step $\Delta t = 1$~h. We denote individual scenarios as  $s\in\mathcal{S}:=\{1,\ldots,|\mathcal{S}|\}$. Each scenario $s$ corresponds to a $T$-length trajectory of  uncertainty realization $\xi_s$ (fixed/non-schedulable loads, reserve, and FRP up/down deployments). 
The finite empirical uncertainty set is denoted by
{$\Xi:=\{\xi_s\in\mathbb{R}^{4T}:s\in\mathcal S\}$}.
Robust feasibility in our formulation therefore means feasibility for every joint trajectory in $\Xi$.

\subsection{Overview of the problem formulation}
The proposed day-ahead co-scheduling problem minimizes the data center's worst-case operating cost over $\Xi$. The cost comprises the real-time energy bill and net revenue from commitments in the reserve and flexible ramping markets, as well as the penalty for delayed and degraded workload execution, and BESS cycling cost.

The operator commits the compute controls (flexible workload schedule and DVFS settings) and BESS controls (charging/discharging and ancillary service capacity offers) in day-ahead before the scenario is realized. For each $\xi_s\in\Xi$, the real-time deployment induces the actual net withdrawal, BESS state of charge (SOC), energy settlement, and cycling cost. The resulting static robust co-optimization problem is:

{\small
  \begin{subequations}
    \label{eq:full-problem}
    \begin{align}
      \min \;&
               \begin{aligned}[t]
                 &\text{workload-quality penalty}~\cref{eq:obj-affine}- \text{capacity revenue}~\cref{eq:ro}\\
                 & \quad + \max_{s\in\mathcal{S}} \bigl[\text{energy cost}~\cref{eq:ro-Renergy}+ \text{BESS cycling cost}~\cref{eq:degradation-cost}\bigr]
               \end{aligned}
               \label{eq:full-problem-obj}\\
      \st\quad
             & \text{system power balance {\cref{eq:ro_grid_loads},}}
               \label{eq:full-problem-balance}\\
             & \text{grid interconnection constraints {\cref{eq:ro_grid_constraints},}}
               \label{eq:full-problem-grid}\\
             & \text{BESS constraints {\cref{eq:B_energy_balance,eq:B_binary,eq:B_charge_limit,eq:B_discharge_limit,eq:B_SOC_bounds,eq:throughput,eq:hard-cycle-limit,eq:degradation-cost},}}
               \label{eq:full-problem-bess}\\
               & \text{compute load constraints {\cref{eq:power-balance-mod,eq:dc-capacity,eq:var-load,eq:rate-bnd-affine,eq:release-affine,eq:work-complete-mod,eq:dvfs-select,eq:obj-affine}.}}
               \label{eq:full-problem-compute}
    \end{align}
  \end{subequations}
}
We solve \cref{eq:full-problem} by enumerating $\Xi$ through its index set $\mathcal S$. The variables in \cref{eq:full-problem} are grouped as:
\begin{itemize}[noitemsep,topsep=0pt,leftmargin=*]
\item \emph{day-ahead controls}: workload schedule, DVFS settings, base BESS dispatch, and ancillary service capacity offers, all common to every $\xi_s\in\Xi$;
\item {\emph{induced scenario quantities}: fixed and total compute load, net grid withdrawal, BESS SOC and throughput, energy settlement, and excess-throughput slack, per $\xi_s$.}
\end{itemize}

\begin{figure*}[t]
 \centering
 \vspace{-2em}
 \includegraphics[width=0.8\textwidth]{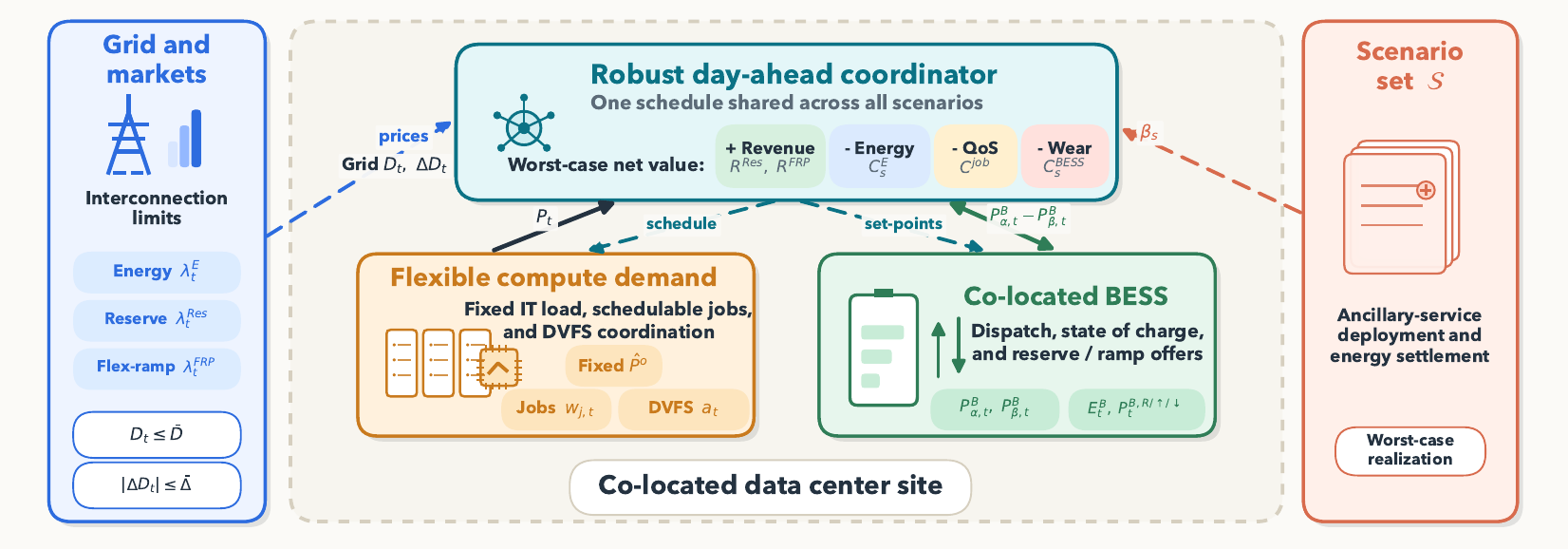}
 \vspace{-1em}
 \caption{Overview of the robust day-ahead co-optimization framework for a data center with co-located BESS. The coordinator jointly schedules flexible compute load and battery operation using market prices and the finite uncertainty set $\Xi$, while enforcing grid-facing load and ramp limits for every $\xi_s\in\Xi$.}
 \vspace{-1em}
 \label{fig:problem-framework-overview}
\end{figure*}

To expand \cref{eq:full-problem}, the remaining subsections detail the grid interface model (\Cref{sec:grid interface}), ancillary service capacity model (\Cref{sec:capacity-revenue}), energy cost (\Cref{sec:energy-cost}), compute load model (\Cref{sec:scheduling_frameworks}), workload-quality penalty (\Cref{sec:workload-penalty}), BESS model (\Cref{sec:bess_model}), and cycling cost (\Cref{sec:cycling-cost}). The fully substituted formulation and nomenclature are given in \cref{app:explicit-formulation}. A high-level overview of the proposed framework is illustrated in \cref{fig:problem-framework-overview}.

\begin{remark}[Static robustness and non-anticipativity]
All controllable decisions in \cref{eq:full-problem} are selected before uncertainty is revealed. Relative to a multistage policy that re-optimizes as information arrives, this static structure is conservative but avoids a scenario tree whose size grows exponentially with the horizon. In practice, lower-level real-time controls can provide additional recourse outside the day-ahead model.
\end{remark}

\subsection{Grid interface constraints}
\label{sec:grid interface}

The grid interface couples the compute load and BESS dispatch through the scenario-dependent net withdrawal
\begin{equation}
D_{t,s} = P^{\mathrm{B}}_{\alpha,t} - P^{\mathrm{B}}_{\beta,t} + P^{\mathrm{DC}}_{t,s}, \quad \forall t\in\mathcal{T},\, s\in\mathcal{S},
\label{eq:ro_grid_loads}
\end{equation}
where $D_{t,s}$ is the grid withdrawal, $P^{\mathrm{DC}}_{t,s}$ is the compute load, and $P^{\mathrm{B}}_{\alpha,t}$ and $P^{\mathrm{B}}_{\beta,t}$ are the BESS base charging and discharging powers, respectively. The BESS base dispatch is a scenario-independent day-ahead decision, whereas $P^{\mathrm{DC}}_{t,s}$ and $D_{t,s}$ depend on the realized fixed load, as {detailed} later.

Under each scenario, the net withdrawal $D_{t,s}$ must satisfy the time-invariant peak demand limit $\bar D$ and symmetric ramp rate limit $\bar\Delta$ as per the interconnection agreement:
\begin{subequations}\label{eq:ro_grid_constraints}
  \begin{align}
    D_{t,s} &\le \bar{D}, \quad \forall t\in\mathcal{T},\, s\in\mathcal{S},
    \label{eq:ro_Cload}\\
    -\bar{\Delta} \le D_{t,s}-D_{t-1,s} &\le \bar{\Delta}, \quad \forall t\in\mathcal{T}\setminus\{1\},\, s\in\mathcal{S},
    \label{eq:ro_Cramp}
  \end{align}
\end{subequations}
where \cref{eq:ro_Cload} and \cref{eq:ro_Cramp} are the peak-load and ramp-rate constraints, respectively.

\subsection{Capacity revenue}
\label{sec:capacity-revenue}
Each ancillary service product is compensated through a capacity payment for committed availability. The day-ahead capacity revenue is
\begin{equation}
R^{\mathrm{cap}} := R^{Res} + R^{\uparrow\downarrow},
\label{eq:ro}
\end{equation}
decomposed into reserve and flexible ramping components
\begin{align}
R^{Res} &:= \sum_{t\in\mathcal{T}} \lambda^{Res}_t\, P^{\mathrm{B},Res}_t\, \Delta t,
\label{eq:ro-Rreserve}\\
R^{\uparrow\downarrow} &:= \sum_{t\in\mathcal{T}} \bigl(\lambda^{\uparrow}_t P^{\mathrm{B},\uparrow}_t + \lambda^{\downarrow}_t P^{\mathrm{B},\downarrow}_t\bigr)\Delta t.
\label{eq:ro-Rramp}
\end{align}
Here $\lambda^{Res}_t, \lambda^{\uparrow}_t, \lambda^{\downarrow}_t$ are the reserve and flexible ramping market-clearing prices, and $P^{\mathrm{B},Res}_t, P^{\mathrm{B},\uparrow}_t, P^{\mathrm{B},\downarrow}_t$ are the BESS reserve and flexible ramping capacity offers (introduced in \Cref{sec:bess_model}). Note that $R^{\mathrm{cap}}$ is locked in by the day-ahead award and independent of the realized scenario. We now present {the must-offer requirement on ancillary commitments, which we add to problem \cref{eq:full-problem} in a subset of our studies}.

\subsubsection{{Minimum ancillary service commitment}}
\label{sec:min-ancillary}
{A must-offer requirement is modeled as a lower bound on the day-ahead capacity offered during a prescribed commitment window denoted by $\mathcal{T}^{\mathrm{AS}}\subseteq\mathcal{T}$. We allow both non-negative product-specific requirements, parameterized by $\underline{P}^{\mathrm{B},Res}_t$, $\underline{P}^{\mathrm{B},\uparrow}_t$, and $\underline{P}^{\mathrm{B},\downarrow}_t$, and an aggregate requirement $\underline{P}^{\mathrm{B},AS}_t$ that permits the optimizer to select the service composition:}
\begin{subequations}
\label{eq:min-ancillary}
\begin{align}
P^{\mathrm{B},Res}_t &\ge \underline{P}^{\mathrm{B},Res}_t,
&& \forall t\in\mathcal{T}^{\mathrm{AS}}, \\
P^{\mathrm{B},\uparrow}_t &\ge \underline{P}^{\mathrm{B},\uparrow}_t,
&& \forall t\in\mathcal{T}^{\mathrm{AS}}, \\
P^{\mathrm{B},\downarrow}_t &\ge \underline{P}^{\mathrm{B},\downarrow}_t,
&& \forall t\in\mathcal{T}^{\mathrm{AS}}, \\
P^{\mathrm{B},Res}_t+P^{\mathrm{B},\uparrow}_t+P^{\mathrm{B},\downarrow}_t
&\ge \underline{P}^{\mathrm{B},AS}_t,
&& \forall t\in\mathcal{T}^{\mathrm{AS}}.
\end{align}
\end{subequations}
Setting unused bounds to zero recovers a specific obligation; setting all bounds to zero removes commitment constraints.

\subsection{Energy cost}
\label{sec:energy-cost}
The total real-time energy cost is
\begin{equation}
\begin{aligned}
C^E_s := \sum_{t\in\mathcal{T}} \lambda^E_t \Delta t \Bigl[&D_{t,s} - \phi^{Res}_{t,s} P^{\mathrm{B},Res}_t \\
&+ \phi^{\uparrow}_{t,s} P^{\mathrm{B},\uparrow}_t - \phi^{\downarrow}_{t,s} P^{\mathrm{B},\downarrow}_t\Bigr],
\end{aligned}
\label{eq:ro-Renergy}
\end{equation}
where $\lambda^E_t$ is the energy price; $\phi^{Res}_{t,s},\phi^{\uparrow}_{t,s}, \phi^{\downarrow}_{t,s}$ represent the fractions of committed reserve, upward and downward flexible ramping capacities respectively, that are deployed in scenario $s$ at time $t$. Note that the real-time energy cost \cref{eq:ro-Renergy} is scenario-dependent. We describe the BESS model next.

\subsection{BESS constraints}
\label{sec:bess_model}
This subsection models the co-located BESS: its intertemporal dynamics and charging/discharging operating limits under flexible load and ancillary service deployment.

\subsubsection{Energy balance and limits}
$\forall t\in\mathcal{T},\, s\in\mathcal{S}$, the intertemporal energy balance is
\begin{align}
E^{\mathrm{B}}_{s,t} = &E^{\mathrm{B}}_{s,t-1}\, + (P^{\mathrm{B}}_{\alpha,t}+\phi^{\uparrow}_{t,s} P^{\mathrm{B},\uparrow}_t)\, \eta^{\mathrm{B}}_{\alpha}\, \Delta t \nonumber\\
&~- (P^{\mathrm{B}}_{\beta,t}+ \phi^{\downarrow}_{t,s} P^{\mathrm{B},\downarrow}_t + \phi^{Res}_{t,s} P^{\mathrm{B},Res}_t)\, \Delta t / \eta^{\mathrm{B}}_{\beta},
\label{eq:B_energy_balance}\\
\underline{E}^{\mathrm{B}} \le& E^{\mathrm{B}}_{s,t} \le \overline{E}^{\mathrm{B}},
\label{eq:B_SOC_bounds}
\end{align}
where $E^{\mathrm{B}}_{s,t}$ is the stored energy at time $t$ in scenario $s$ and $\eta^{\mathrm{B}}_{\alpha},\eta^{\mathrm{B}}_{\beta}$ are the charging and discharging efficiencies. The deployment factors $\phi^{Res}_{t,s},\phi^{\uparrow}_{t,s},\phi^{\downarrow}_{t,s}$ are the same as in \cref{eq:ro-Renergy}. Note that each scenario's realized ancillary service deployment propagates through \cref{eq:B_energy_balance} into a distinct trajectory for $E^{\mathrm{B}}_{s,t}$. $\underline{E}^{\mathrm{B}}, \overline{E}^{\mathrm{B}}$ are the lower and upper admissible stored-energy limits in \cref{eq:B_SOC_bounds}.
\subsubsection{Operating mode and charging limits}
The battery operates in charging or discharging mode, but not both simultaneously:
\begin{equation}
0 \le z^{\alpha}_{t} + z^{\beta}_{t} \le 1, \quad z^{\alpha}_{t}, z^{\beta}_{t} \in \{0,1\}, \;\forall t\in\mathcal{T},
\label{eq:B_binary}
\end{equation}
where $z^{\alpha}_{t},z^{\beta}_{t}$ are charging and discharging  indicators, respectively. For $z^{\alpha}_{t}=1$, the base charging and upward-ramping capacity are capped by the maximum charging rate:
\begin{equation}
{P^{\mathrm{B}}_{\alpha,t} \ge 0,P^{\mathrm{B},\uparrow}_{t} \ge 0,} P^{\mathrm{B}}_{\alpha,t} + P^{\mathrm{B},\uparrow}_{t} \le z^{\alpha}_{t} P^{\mathrm{B},\max}_{\alpha}, \;\forall t\in\mathcal{T},
\label{eq:B_charge_limit}
\end{equation}
where $P^{\mathrm{B},\max}_{\alpha}$ is the maximum charging power. When $z^{\beta}_{t}=1$, the base discharging, reserve and downward-ramping capacities are capped by the maximum discharging rate:
\begin{equation}
\begin{aligned}
&{P^{\mathrm{B}}_{\beta,t} \ge 0,\quad P^{\mathrm{B},Res}_{t} \ge 0,\quad P^{\mathrm{B},\downarrow}_{t} \ge 0,}\\
&P^{\mathrm{B}}_{\beta,t} + P^{\mathrm{B},Res}_{t} + P^{\mathrm{B},\downarrow}_{t} \le z^{\beta}_{t} P^{\mathrm{B},\max}_{\beta}, \;\forall t\in\mathcal{T},
\end{aligned}
\label{eq:B_discharge_limit}
\end{equation}
where $P^{\mathrm{B},\max}_{\beta}$ is the maximum discharging power. 

\subsubsection{Cycle warranty and cost}
\label{sec:cycling-cost}
The total energy throughput $E^{\text{th}}_{s}$ in scenario $s\in \mathcal{S}$ accumulates all charge and discharge flows:
\begin{equation}
\begin{aligned}
E^{\text{th}}_{s} = \sum_{t \in \mathcal{T}} \Delta t\, \bigl(&P^{\mathrm{B}}_{\alpha,t} + \phi^{\uparrow}_{t,s} P^{\mathrm{B},\uparrow}_t \\
&+ P^{\mathrm{B}}_{\beta,t} + \phi^{Res}_{t,s} P^{\mathrm{B},Res}_t + \phi^{\downarrow}_{t,s} P^{\mathrm{B},\downarrow}_t\bigr){.}
\end{aligned}
\label{eq:throughput}
\end{equation}
The manufacturer-specified cycle warranty $\overline{E}^{\text{th}}$\footnote{{Nominal} warranty throughput budget is equal to twice the BESS energy capacity times the preferred daily cycle limit (e.g., 0.5~cycles/day~\cite{balducci2019nantucket})} is enforced as a soft constraint on $E^{\text{th}}_{s}$ with a scenario-indexed slack $\widetilde{E}^{\text{th}}_s\ge 0$:
\begin{equation}
E^{\text{th}}_{s} \le \overline{E}^{\text{th}} + \widetilde{E}^{\text{th}}_s, \;\forall s\in\mathcal{S}{.}
\label{eq:hard-cycle-limit}
\end{equation}
The BESS cycling cost is hence enforced on $E^{\text{th}}_{s}$ as:
\begin{equation}
C^{BESS}_s := c^{\text{deg}}\, \widetilde{E}^{\text{th}}_s,
\label{eq:degradation-cost}
\end{equation}
where $c^{\text{deg}}$ is the marginal degradation cost per unit excess throughput; a natural choice is to set $c^{\text{deg}}$ to the capital cost of installing a new BESS amortized over its rated lifetime.

To complete our modeling section, we detail costs, controls and constraints of the compute load next.

\subsection{Compute load constraints}
\label{sec:scheduling_frameworks}
The compute load consists of two job categories:
\begin{itemize}[noitemsep,topsep=0pt,leftmargin=*]
\item \textbf{Fixed/non-schedulable jobs} that are always-on workloads whose execution timing is not negotiable, such as web serving, real-time inference, databases, and monitoring. 
\item \textbf{Schedulable jobs} that are delay-tolerant workloads that must be completed within release-deadline windows but admit flexibility in when and at what rate they execute, such as batch analytics, model training, and batch inference.  The aggregate job portfolio used in the case studies is given in \cref{tab:job_portfolio}, \cref{app:job-portfolio}.
\end{itemize}

The operator shapes these loads through two control knobs detailed below: job level workload scheduling within release-deadline windows~\cite{hall2025carbon} and a system-level dynamic voltage and frequency scaling (DVFS) adjustment~\cite{piga2024expanding}. We treat both as deterministic day-ahead decisions. In practice, residual stochasticity in job runtimes, processor performance, and intra-day arrivals exists, and it is handled by real-time scheduling mechanisms outside the model. 

\subsubsection{Workload power}
The aggregate compute power decomposes into a scenario-dependent fixed-load component $P^{\mathrm{fixed}}_{t,s}$ and a variable (schedulable) component $P^{\mathrm{var}}_t$:
\begin{equation}
P^{\mathrm{DC}}_{t,s} = P^{\mathrm{fixed}}_{t,s} + P^{\mathrm{var}}_t, \quad \forall t\in\mathcal{T},\, s\in\mathcal{S},
\label{eq:power-balance-mod}
\end{equation}
which must respect the internal server hardware power limit:
\begin{equation}
P^{\mathrm{DC}}_{t,s} \le \bar{P}^{\mathrm{DC}}, \quad \forall t\in\mathcal{T},\, s\in\mathcal{S}.
\label{eq:dc-capacity}
\end{equation}
Here, the internal server hardware power limit $\bar{P}^{\mathrm{DC}}$ constrains server-side compute power and is distinct from the grid side interconnection limit $\bar{D}$ in \cref{eq:ro_Cload} that constrains the net grid withdrawal after including BESS injections.\\
The fixed job power is treated as inelastic and DVFS-independent. It always executes in full at its forecast power: 
\begin{equation}
    P^{\mathrm{fixed}}_{t,s}=\widehat P^{\mathrm{fixed}}_t+\varepsilon_{t,s},
\end{equation}
that combines a trace-based point forecast with the additive forecast error in $\xi_s\in\Xi$. 
\subsubsection{Schedulable load}
\label{sec:schedulable-jobs}
The schedulable job power $P^{\mathrm{var}}_t$ is the sum of per-job DVFS-weighted execution rates:
\begin{equation}
P^{\mathrm{var}}_t = \sum_{j\in\mathcal{J}} P^{\mathrm{var}}_{j,t}, \quad \forall t\in\mathcal{T},
\label{eq:var-load}
\end{equation}
where $P^{\mathrm{var}}_{j,t}\ge 0$ is the DVFS-weighted execution rate of job $j$ at time $t$ (introduced in \Cref{sec:dvfs} below). The execution rate for schedulable job $j\in\mathcal{J}$ with release time $S_j$ and deadline $d_j$ is subject to multiple constraints. First, no work is delivered before the release time $S_j$:
\begin{equation}
P^{\mathrm{var}}_{j,t}=0, \quad\forall j\in\mathcal{J},\, t<S_j.
\label{eq:release-affine}
\end{equation}
Second, the cumulative work delivered to each job must meet its total work requirement $W_j$ up to a non-negative slack $\widetilde{W}_j$:
\begin{equation}
W_j-\widetilde{W}_j \le \sum_{t\in\mathcal{T}} P^{\mathrm{var}}_{j,t}\,\Delta t \le W_j, \quad\forall j\in\mathcal{J}.
\label{eq:work-complete-mod}
\end{equation}
Post-deadline work (beyond the deadline $d_j$) is additionally penalized in \cref{eq:obj-affine}. The final set of constraints for $P^{\mathrm{var}}_t$ arises from DVFS, as detailed next.
\subsubsection{Discrete dynamic voltage and frequency scaling (DVFS)}
\label{sec:dvfs}
Each active schedulable job selects one DVFS level per period from a discrete set of hardware-realizable performance states (e.g., ACPI P-states~\cite{guenter2011managing}). Let $\mathcal{L} = \{0, 1, \ldots, L\}$ index the levels, with $\ell = 0$ designating the nominal level. For each job $j\in\mathcal{J}$ and period $t\in\mathcal{T}$, at most one level is selected using
\begin{equation}
z_{j,t,\ell} \in \{0,1\},\;\forall \ell\in\mathcal{L}, \sum_{\ell\in\mathcal{L}} z_{j,t,\ell} \le 1, \forall j\in\mathcal{J},\, t\in\mathcal{T}.
\label{eq:dvfs-select}
\end{equation}
Each level $\ell$ is characterized by a per-job rate cap $\overline{P}^{\mathrm{var}}_{j,\ell}$. The execution rate $P^{\mathrm{var}}_{j,t}$ is thus bounded as:
\begin{equation}
0 \le P^{\mathrm{var}}_{j,t} \le \sum_{\ell\in\mathcal{L}} \overline{P}^{\mathrm{var}}_{j,\ell}\, z_{j,t,\ell}, \quad\forall j\in\mathcal{J},\, t\in\mathcal{T}.
\label{eq:rate-bnd-affine}
\end{equation}
Note that higher-frequency levels admit higher per-slot service and power, while lower-frequency levels tighten the cap.

\subsection{Workload-quality penalty}
\label{sec:workload-penalty}
The workload-quality penalty has three components and is given by:
\begin{equation}
C^{\mathrm{job}} := C^{\mathrm{DVFS}} + C^{\mathrm{unf}} + C^{\mathrm{tardy}},
\label{eq:obj-affine}
\end{equation}
with each component defined as follows.\\
$C^{\mathrm{DVFS}}$ is the DVFS-deviation penalty and is charged for each job at each period based on the DVFS level selected:
\begin{equation}
C^{\mathrm{DVFS}} := \sum_{j\in\mathcal{J}}\sum_{t\in \mathcal{T}}\sum_{\ell\in\mathcal{L}} c^{\mathrm{DVFS}}_{j,\ell}\, z_{j,t,\ell}.
\label{eq:cost-dvfs}
\end{equation}
Here $c^{\mathrm{DVFS}}_{j,\ell}\ge 0$ is the per-level cost of operating job $j$ at DVFS level $\ell$, with $c^{\mathrm{DVFS}}_{j,0}=0$ at the nominal level. $C^{\mathrm{DVFS}}$ discourages aggressive DVFS variations that can hinder performance and cause operational problems in control-system stability and thermal capacity~\cite{patel2024characterizing,stojkovic2025tapas}. \\
$C^{\mathrm{unf}}$ is the unfinished-work penalty for jobs that do not complete within the horizon and is given by:
\begin{equation}
C^{\mathrm{unf}} := \sum_{j\in \mathcal{J}} c^{\mathrm{unf}}_j\, \widetilde{W}_j,
\label{eq:cost-unf}
\end{equation}
where $c^{\mathrm{unf}}_j > 0$ is the job-specific per unit \emph{value of lost compute load} that the {data center} operator foregoes when job $j$ fails to complete within the scheduling horizon.\\
Finally, $C^{\mathrm{tardy}}$ is the tardy-execution penalty
\begin{equation}
C^{\mathrm{tardy}} := \sum_{j\in \mathcal{J}}\sum_{\substack{t\in \mathcal{T}\\ t>d_j}} c^{\mathrm{tardy}}_j\, (t-d_j)\, P^{\mathrm{var}}_{j,t}\, \Delta t,
\label{eq:cost-tardy}
\end{equation}
where $c^{\mathrm{tardy}}_j > 0$ is the per-job, per-unit cost of post-deadline execution, weighted by delay $(t-d_j)$. 

The complete mathematical formulation after inserting all constraints and costs into the robust problem \cref{eq:full-problem} is given in \cref{app:explicit-formulation}. The resulting finite-scenario formulation is a mixed-integer linear program (MILP): the maximum over scenarios admits a linear epigraph representation, while the BESS charging/discharging mode indicators and discrete DVFS-level selectors are binary in \cref{eq:B_binary,eq:dvfs-select}, respectively. In Appendix~\ref{app:lp-relaxation}, we assess the direct LP relaxation of both binary sets over the tested instances, which yields a mean objective gap of 0.060\% and a mean computational speedup of 18.4$\times$. Before analyzing the computational results for our model in the next section, we provide the following elucidatory remarks on our modeling choices.

\begin{remark}[DVFS-level granularity]
\label{rem:dvfs-integrality}
The DVFS selector $z_{j,t,\ell}$ is modeled as binary in \cref{eq:dvfs-select}, so that all servers assigned to job $j$ in period $t$ run at a single DVFS level $\ell$. If the {data center} operator is instead willing to assign distinct DVFS frequencies to individual servers within a job's allocation, the integrality constraint $z_{j,t,\ell}\in\{0,1\}$ may be relaxed to $z_{j,t,\ell}\in[0,1]$, in which case $z_{j,t,\ell}$ can be interpreted as the fraction of servers running job $j$ at level $\ell$ in period $t$.
\end{remark}

\begin{remark}[Service-quality hierarchy in penalty]\label{rem:service-quality}
The coefficients of penalty terms in \cref{eq:obj-affine} are chosen to encode the priority order $c^{\mathrm{unf}}_j > c^{\mathrm{tardy}}_j \gg c^{\mathrm{DVFS}}_{j,\ell}$ after accounting for the units and typical magnitudes. This hierarchy treats non-completion as the least desirable outcome, permits deadline violation only when needed, and allows DVFS adjustment as the lowest-cost flexibility lever. Using publicly available AWS pricing and NVIDIA system-power data, we obtain an illustrative revenue intensity of approximately 3400~\$/MWh of IT energy~\cite{aws2026p5pricing,nvidia2023dgxh100}. Based on this benchmark, we set the base unfinished-work penalty to 2000~\$/MWh and apply service-priority multipliers to obtain a range of 200--5000~\$/MWh for our schedulable jobs, described in \cref{tab:job_portfolio}. Further, we set the tardiness penalty to 5\% of the corresponding unfinished-work penalty per hour. Note that these values are case-specific calibrations rather than universal estimates.
\end{remark}

\section{Case Studies}
\label{sec:case_studies}
We evaluate the proposed framework on a synthetic data center setup with 100~MW load capacity, hourly Google compute load trace data~\cite{googletrace2020}, and energy, reserve, and flexible ramping prices from CAISO~\cite{caiso2025ramp} and PJM~\cite{pjm2025reserve}. Unless otherwise stated, the horizon is 24 hourly intervals.\footnote{Code and examples are available at \url{https://github.com/ShaohuiLiu/DCFlex}.}

\subsection{Data and setup}
\label{sec:data_setup}
The base co-located battery is rated at 36~MWh/12~MW. The representative dispatch uses the 100~MW load-cap and ramp rate of 15~MW/h. Subsequently, an interconnection-sensitivity study sweeps around the base point $(\bar D,\bar \Delta)=(105~\mathrm{MW},15~\mathrm{MW/h})$. The 31 daily instances use consecutive 24-hour slices of the scaled Google trace and reproducible, day-specific market profiles. Rather than replaying a single market day, we construct representative hourly prices from historical CAISO/PJM ranges and diurnal patterns and add bounded random perturbations to capture day-to-day variation. The scenario set $\Xi$ contains 50 joint in-sample trajectories of ancillary service deployment and fixed-load forecast error. Historical CAISO FRP binding frequencies and hourly profiles and PJM reserve statistics parameterize the deployment factors, which are restricted to $[0,1]$. The fixed-load forecast error $\varepsilon_{t,s}$ is sampled independently from a zero-mean Gaussian distribution with a standard deviation of 2.5~MW, approximately 3\% of the representative day's 81.0--85.1~MW fixed-load level. The total workload from schedulable jobs is 340~MWh, with an hourly average of roughly 14~MWh, as detailed in \cref{app:job-portfolio}. The job-specific workload-quality coefficients are also reported in \cref{tab:job_portfolio} and follow Remark \ref{rem:service-quality}. The per-level DVFS-deviation penalties are set proportional to the relative frequency offset of each level, with maximum-deviation cost $c^{\mathrm{DVFS}}_{j,\ell}=1000$~\$/job-period and $c^{\mathrm{DVFS}}_{j,0}=0$ at the nominal level. The battery degradation coefficient is $c^{\mathrm{deg}}=45$~\$/MWh.

As the system scale, market data, and service-quality coefficients are calibrated for representative cases, the numerical costs and penalties should be interpreted for relative comparisons and marginal values for {data center} flexibility sources, under different interconnection and workload conditions.

\begin{figure*}[t]
\centering
\vspace{-2em}
\includegraphics[width=0.48\textwidth]{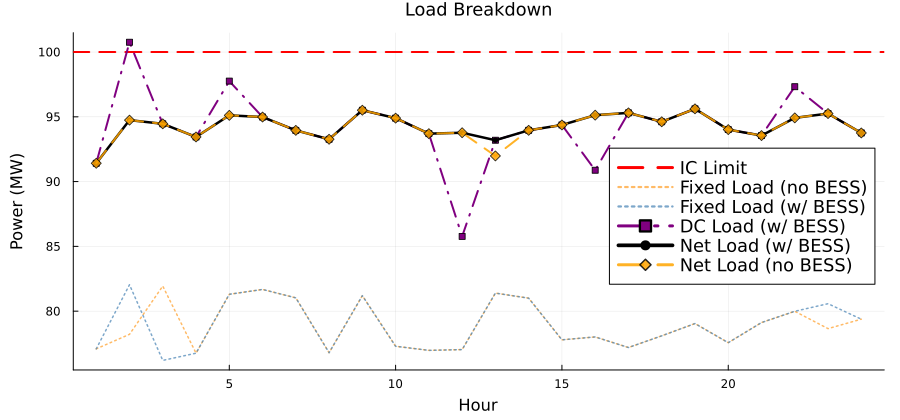}%
\hfill
\includegraphics[width=0.48\textwidth]{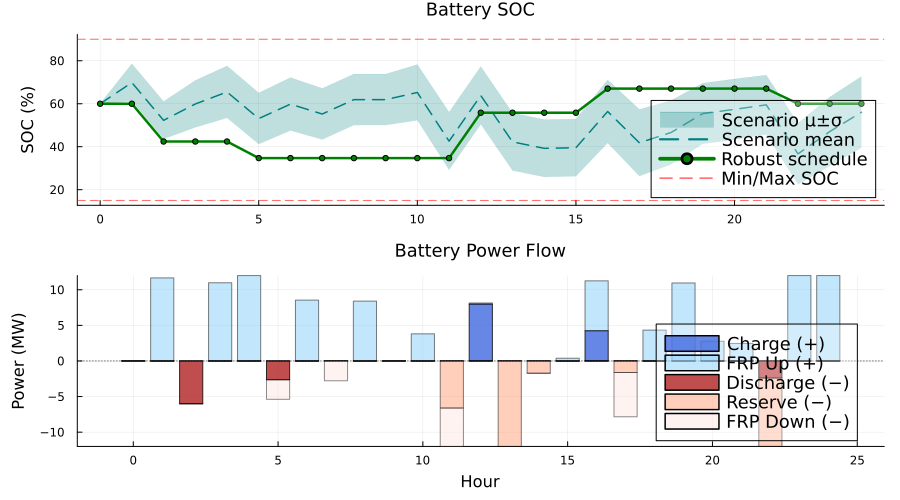}%
\vspace{-1em}
\caption{Representative daily dispatch for a 100~MW data center with a 36~MWh/12~MW co-located BESS; both panels report the same optimized instance. Left: fixed compute, internal, and net load with and without BESS. Upper right: solid green denotes the base-schedule state of charge (SOC) before ancillary service deployment; dashed teal and shading denote the pointwise scenario mean and mean $\pm$ one standard deviation, respectively. Lower right: day-ahead base charge/discharge and committed ancillary service capacities.}
\vspace{-1em}
\label{fig:dispatch_main}
\end{figure*}

\subsection{Representative Daily Dispatch}
\label{sec:rep_daily}
We first illustrate how our proposed framework's three control knobs (workload scheduling, DVFS, and BESS dispatch) jointly use flexibility to respect interconnection constraints while reducing the daily operating cost. We solve the robust day-ahead MILP \cref{eq:full-problem} for a representative daily instance for the base setting, with the BESS having a 60\% initial and terminal SOC requirement. The left panel of \Cref{fig:dispatch_main} shows that with BESS, the internal data center load reaches beyond the 100~MW interconnection limit in selected hours because the battery offsets those excursions, keeping the grid-facing net load below 95.62~MW. The no-BESS case has no behind-the-meter buffer, so its internal data center load and grid-facing net load coincide. Both solutions complete the schedulable workload without peak load, ramp, or quality slack; the BESS reduces the daily operating cost from \$228{,}418 to \$223{,}530, a modest cost reduction of \$4{,}888. 

The right panel of \Cref{fig:dispatch_main} presents the BESS SOC trajectories. The base dispatch charges 12.3~MWh and discharges 11.1~MWh while committing 31.5~MW-h of reserve, 107.5~MW-h of up-ramping capacity, and 17.1~MW-h of down-ramping capacity. Note that the optimization enforces the SOC limits separately for every trajectory in $\Xi$. Relative to the no-BESS case, the per-job DVFS level selections $z_{j,t,\ell}$ differ in a small number of hours, indicating that workload scheduling and DVFS {act} in coordination with the BESS.

To evaluate the value of flexible workloads, BESS and DVFS under stressed cases, we now analyze scenarios where interconnection limits are binding. For {this} set of results in {Sections} \ref{sec:interconnection}{--}\ref{sec:dvfs_case}, we solve 31 daily instances of the robust co-optimization for the 100-MW data center. Each daily instance in the 31 days uses a different 24-hour real base-load trace. 
\subsection{Interconnection Sensitivity Studies}
\label{sec:interconnection}
\begin{figure}[t]
\centering
\vspace{-1.5em}
\includegraphics[width=\columnwidth]{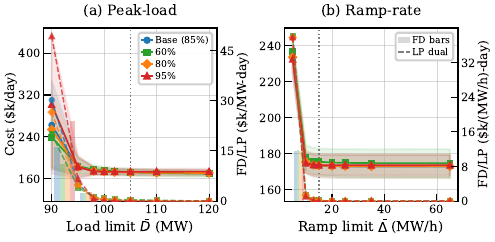}
\vspace{-2em}
\caption{Interconnection sensitivity under alternative load compositions over 31 daily instances. The solid curves and shaded bands denote the mean and standard deviation of the MILP daily operating cost, where lower values are better. Colors correspond to the different composition of fixed loads, base case with approximately 85\% fixed load, and counterfactual fixed-load shares of 60\%, 80\%, and 95\%. On the secondary axis, the semi-transparent bars show daily finite-difference (FD) marginal values and the dashed curves show the mean daily aggregate LP duals for the same composition. (a) Composition matters mainly when the peak-load limit is binding: the high-fixed-load 95\% case exhibits the largest binding-constraint cost under tight limits, while marginal values nearly vanish once the limit reaches about 105~MW. \revise{(b) Under the ramp-limit sweep, the operating-cost and marginal-value curves are nearly composition-invariant, indicating that ramp-limit binding is driven primarily by the aggregate net load trajectory rather than the flexible-load share.} The dotted vertical line in each panel marks the base case $(\bar D,\bar \Delta)=(105~\mathrm{MW},15~\mathrm{MW/h})$.}
\vspace{-2em}
\label{fig:ic_sensitivity}
\end{figure}
We next quantify how the economic value of coordinated flexibility depends on the available interconnection capacity and on the composition of fixed versus schedulable load. Starting from the base case $(\bar D,\bar \Delta)=(105~\mathrm{MW},15~\mathrm{MW/h})$ with a 36~MWh/12~MW BESS, we sweep both the peak-load limit and the ramp rate limit and solve the MILP \cref{eq:full-problem} across the 31 daily instances. In addition to the natural base composition generated from the real-market case, which has approximately 85\% fixed load on average, we construct counterfactual portfolios in which fixed jobs account for 60\%, 80\%, or 95\% of total compute workload, with the remainder treated as schedulable workloads. \revise{At each sweep point, the finite-difference (FD) sensitivity of the MILP captures the impact of directly relaxing the interconnection limits on the average daily operating cost.} We also fix the MILP integer variables at each $(\bar D,\bar \Delta)$ and re-solve the resulting LP to obtain local shadow prices (dual variables) for the realized discrete operating mode. To compare those LP duals with the FD sensitivities, we aggregate the hourly duals within each day, summing $\lambda^{\mathrm{load}}_t$ for the load-limit sweep and $\lambda^{\mathrm{ramp\text{-}up}}_t+\lambda^{\mathrm{ramp\text{-}down}}_t$ for the ramp-limit sweep. We then average the daily aggregates across the 31 days. 

\Cref{fig:ic_sensitivity} reports the resulting daily operating costs and marginal-value measures.
In \Cref{fig:ic_sensitivity}(a), once the load limit reaches 105~MW, the operating cost stabilizes near \$172k/day for all workload compositions, and the corresponding LP duals are below about \$0.5k/MW-day, indicating little effect of job composition on the marginal value of additional load headroom. Under the tight 90~MW load limit, the mean operating cost ranges from \$239.4k/day for the 60\% fixed-load case to \$302.0k/day for the 95\% fixed-load case, while the daily aggregate LP dual rises from about \$20.0k/MW-day to about \$49.3k/MW-day. Compared to the operating cost at the 105~MW limit, this represents an increase in operating cost by almost 40\% even for the most flexible workload composition. More crucially, across workload compositions, increasing the fixed-load share from 60\% to 95\% increases the operating cost by more than 26\%, and more than doubles the shadow price of the peak-load limit.
The results reveal that workload composition matters mainly when the interconnection constraints are binding, especially for the peak-load limit. 

In contrast, \Cref{fig:ic_sensitivity}(b) shows that the ramp-limit sensitivity is nearly composition-invariant. While the cost increases by almost 40\% when the ramp limit is tightened to 5~MW/h from 15~MW/h, the mean spread is only about \$4.3k/day across the 60\%, 80\%, and 95\% fixed-load cases. Similarly, the daily aggregate LP duals remain almost identical at roughly \$37.7k--\$38.0k/(MW/h)-day. Over the base-relevant 5--10~MW/h interval, the finite-difference marginal value stays near \$11.6k/(MW/h)-day for all three compositions. 

\begin{figure}[ht]
\centering
\includegraphics[width=.75\columnwidth]{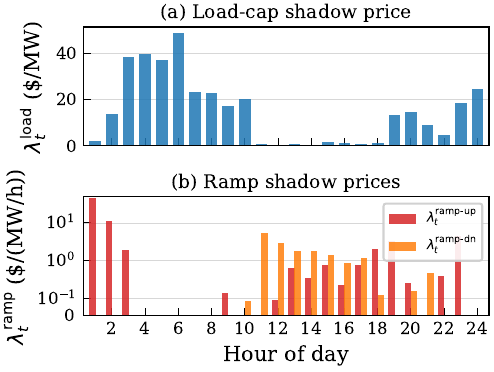}
\vspace{-1em}
\caption{Hourly LP shadow prices for the base interconnection case over 31 daily instances. Panel~(a) reports the hourly mean of $\lambda^{\mathrm{load}}_t$. Panel~(b) reports the hourly mean of $\lambda^{\mathrm{ramp\text{-}up}}_t$ and $\lambda^{\mathrm{ramp\text{-}down}}_t$ on a symmetric-log scale.}
\vspace{-1em}
\label{fig:lp_dual_temporal}
\end{figure}
The distribution of the realized hourly LP duals of constraints \cref{eq:ro_grid_constraints} under the base composition of fixed and flexible loads at interconnection limits $(\bar D,\bar \Delta)=(105~\mathrm{MW},15~\mathrm{MW/h})$ is shown in \Cref{fig:lp_dual_temporal}. The shadow price for the load limit \cref{eq:ro_Cload} is concentrated in a small number of hours, with the largest mean hourly value around 49~\$/MW. The shadow prices for the ramp limits \cref{eq:ro_Cramp} are more episodic as observed in  \Cref{fig:lp_dual_temporal}(b), where the shadow prices are plotted on a logarithmic scale. Averaged across 31 days, the upward-ramp dual peaks around 47~\$/MW/h near the beginning of the day, and hovers around or below 1~\$/MW/h for the rest of the day. The downward-ramp dual is much smaller and again concentrated around transition hours. To {complement} this study on {the} impact of workload composition, we next evaluate the value of the BESS (36~MWh/12~MW) by comparing the daily operating cost under stressed interconnection limits with and without storage.

\subsection{Flexibility Value of BESS Under Interconnection Stress}
\begin{figure}[t]
\centering
\vspace{-2em}
\includegraphics[width=\columnwidth]{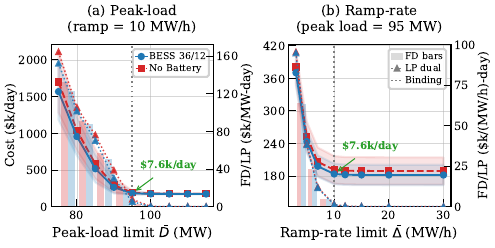}
\vspace{-2em}
\caption{Battery value under stressed interconnection limits over 31 daily instances. The 36~MWh/12~MW BESS has modest value under loose interconnection limits, but its incremental value rises sharply when the peak-load or ramp-rate limit becomes binding. Solid and dashed curves use the left axis; FD bars and dotted LP-dual curves use the right axis in \$k per unit of daily interconnection capacity.}
\vspace{-2em}
\label{fig:stressed_bess_value}
\end{figure}

As shown in \Cref{fig:stressed_bess_value}(a) for base workload composition, under loose peak-load limits above 105~MW, the BESS reduces the operating cost by about \$3.3--\$3.5k/day, with a comparable distribution between its base-case ancillary service and interconnection-flexibility contributions. At the binding peak load of 95~MW, however, the incremental cost reduction rises to about \$7.5k/day, reflecting a gain of greater than 100\%. The cost reduction further increases to about \$129k/day under severe load stress at 75~MW. Additionally, with the peak load limit fixed at 95~MW in \Cref{fig:stressed_bess_value}(b), the BESS cost reduction remains between  \$7.4k/day and \$10.4k/day across ramp limits from 3 to 30~MW/h, while at the 10~MW/h binding point, it reduces the operating-cost standard deviation from about \$25.3k/day to \$19.0k/day. Note that the BESS earns minimal ancillary service revenue across these stressed cases; the additional value under tight limits is primarily operational via buffering net load excursions. 

\subsection{Minimum Market-Participation Sensitivity}
\label{sec:min_participation}
We now evaluate the impact of must-offer ancillary commitments by {adding the constraints in \cref{eq:min-ancillary} to the robust problem \cref{eq:full-problem}} and solving it for 31 daily instances. For the 12~MW/36~MWh BESS, a constant per-period floor is imposed during the eight-hour evening window $\mathcal{T}^{\mathrm{AS}}=\{15,\ldots,22\}$ and swept from 0 to 12~MW. We compare an aggregate floor on reserve plus upward and downward FRP.

The aggregate obligation remains feasible on all 31 days as the floor is {raised} to 10~MW. At 8~MW (67\% of the BESS power rating), it increases mean net operating cost by approximately 0.4\% while maintaining 100\% workload completion; ancillary service revenue increases from \$1.09k/day to \$1.23k/day, whereas BESS cycling decreases from 0.081 to 0.032 equivalent cycles/day. Feasibility declines to 97\% of daily instances at 11~MW and 26\% at 12~MW, indicating that power and energy headroom becomes binding when the must-offer commitment nears the BESS power rating.

The reserve-only obligation is substantially more restrictive. A 1~MW floor increases mean net operating cost by \$0.31k/day; at 3~MW, the increase reaches \$0.80k/day, mean ancillary service revenue decreases from \$1.09k/day to \$0.77k/day, and mean workload completion decreases to 99.9\%. All daily instances are infeasible at floors of 4~MW or greater. Unlike the aggregate requirement, the reserve-only constraint cannot substitute among BESS services. Thus, the cost and feasibility of a must-offer service depend on the aggregate or individual ancillary service product requirement.

\subsection{Discrete DVFS Operating Sets}
\label{sec:dvfs_case}
Finally, we isolate the contribution of processor-level flexibility by comparing no DVFS with four hardware-realizable DVFS operating sets under {the} \((\bar D,\bar \Delta)=(100~\mathrm{MW},15~\mathrm{MW/h})\) load-limit configuration: 3 levels over a \(\pm 5\%\) range, 5 levels over \(\pm 10\%\), 7 levels over \(\pm 15\%\), and 9 levels over \(\pm 20\%\) around the nominal frequency. 

\begin{figure}[t]
\centering
\vspace{-2em}
\includegraphics[width=\columnwidth]{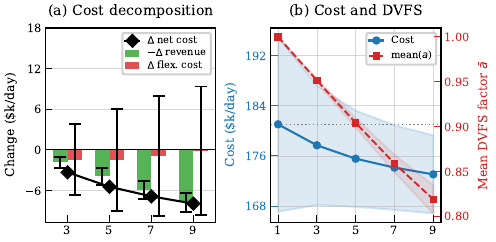}
\vspace{-2em}
\caption{Multiday discrete-DVFS case study under the 100~MW load-limit configuration. The cases compare no DVFS with 3-level \(\pm 5\%\), 5-level \(\pm 10\%\), 7-level \(\pm 15\%\), and 9-level \(\pm 20\%\) operating sets. Overall, allowing DVFS improves the mean daily operating cost and reduces day-to-day variability by providing additional scheduling and interconnection flexibility.}
\label{fig:ablation_dvfs_levels}
\end{figure}

\Cref{fig:ablation_dvfs_levels} shows that enabling DVFS improves average performance relative to the no-DVFS baseline. The no-DVFS case attains a mean operating cost of \$181.0k\(\pm\$13.9\)k, whereas the 9-level case attains \$173.1k\(\pm\$6.2\)k. The 5-, 7-, and 9-level cases reduce the operating cost by \$5.5k/day, \$6.9k/day, and \$8.0k/day, respectively. DVFS also increases the mean completion rate from 97.3\% without DVFS to 100\% in the 9-level case.

To conclude our results section, we next study optimal BESS configuration (sizing and technology) that minimizes total daily cost (operating cost plus amortized capital) for a 100-MW data center under normal operations.

\subsection{Battery Sensitivity Studies}
\label{sec:battery_sensitivity}
This subsection shows how {the optimal} BESS size shifts with cycling-allowance and discount-rate assumptions. We solve 31 daily instances of the robust co-optimization for a 100-MW data center across a range of BESS fleet sizes comprising  $n=1,\ldots,20$ identical BESS units, together with daily cycling limit $\bar N_{\text{cyc}}\in\{0.5,\,1.0,\,1.5,\,2.0\}$~cy/day.

\begin{figure}[ht]
\vspace{-1em}
\centering
{
\includegraphics[width=\columnwidth]{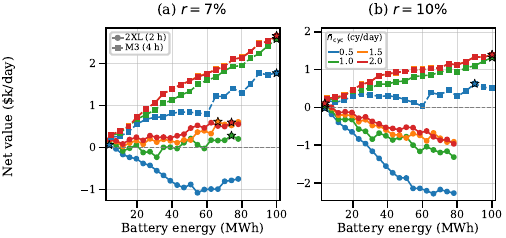}
}
\vspace{-2em}
\caption{Commercial BESS sizing over $D=31$ daily robust instances. Each daily objective is the worst-case operating cost over $\Xi$, and displayed values are arithmetic means of paired same-day comparisons with the no-BESS case. Panels report mean net value versus installed energy at (a) $r{=}7\%$ and (b) $r{=}10\%$. Colors denote cycling limits, solid and dashed curves denote Megapack~2XL and Megapack~3, respectively, and stars identify maximizers over the tested fleet sizes.}
\label{fig:battery_sensitivity_main}
\end{figure}
While the prior experiments in this section consider a benchmark battery rated at 36~MWh/12~MW, the BESS sizing study compares two commercial Tesla Megapack products: the 2-hour Megapack~2XL, which offers higher power but lower energy per unit (1.9~MW/3.9~MWh, \$1.2\,M, $\eta_{\text{RT}}=92\%$), and the 4-hour Megapack~3, which offers lower power but higher energy per unit at slightly lower cost (1.25~MW/5~MWh, \$1.0\,M, $\eta_{\text{RT}}=91\%$). Here $\eta_{\text{RT}}$ is the round-trip efficiency. Each sizing instance initializes the BESS at $e_0{=}60\%$ SOC but, unlike the representative dispatch in \Cref{sec:rep_daily}, does not impose a terminal-SOC constraint.

For each day $d$ and fleet size $n$, $Z_d^{\star}(n)$ denotes the optimized worst-case operating cost over the corresponding 50-trajectory set $\Xi_d$. $Z^{\star}_d(0)$, the no-BESS baseline, has a mean over the 31 daily instances {of} \$176{,}278/day. The \emph{cost reduction} by storage is
$\frac{1}{D}\sum_{d=1}^{D}\bigl[Z^{\star}_d(0)-Z^{\star}_d(n)\bigr]$. 

As the final SOC may fall below initial SOC and  artificially reduce the reported cost reduction, we correct for this \emph{SOC-depletion} artifact in the cost-reduction using:
\begin{equation}
\label{eq:soc_depl}
V^{\mathrm{dep}}(n)=\frac{1}{D}\sum_{d=1}^{D}\max\!\bigl(0,\;e_0-e_{T,d}\bigr)\,n\,E^{\mathrm{cap}}\,\eta^{\mathrm{B}}_{\beta}\,\bar\lambda_{e,d},
\end{equation}
where $e_{T,d}$ is the terminal SOC fraction, $E^{\mathrm{cap}}$ is the per-unit energy capacity, and $\bar\lambda_{e,d}$ is the mean energy price on day $d$. We now present the value addition $\mathrm{VA}(n)$, {which} includes ancillary service (AS) $(R^{Res},R^{\uparrow\downarrow}$) and non-AS residuals $\mathrm{VA}^{\mathrm{NA}}$:\vspace{-3pt}
\begin{align}
\mathrm{VA}(n)&= \frac{1}{D}\sum_{d=1}^{D}\bigl[Z^{\star}_d(0)-Z^{\star}_d(n)\bigr]-V^{\mathrm{dep}}.\label{eq:value_added}\\
&=R^{Res}+R^{\uparrow\downarrow}+\mathrm{VA}^{\mathrm{NA}}
\end{align}
To identify the economically optimal fleet size, we subtract the annualized capital cost of BESS from the \revise{daily cost reduction} to get net value (NV):
\begin{subequations}
\label{eq:net_value}
\begin{align}
\mathrm{NV}(n)&=\mathrm{VA}(n)-\frac{n\,C^{\mathrm{unit}}\cdot\mathrm{CRF}(r,N)}{365}, \\
\mathrm{CRF}&=\frac{r(1{+}r)^N}{(1{+}r)^N{-}1}.
\end{align}
\end{subequations}
The capital recovery factor ($\mathrm{CRF}$) annualizes the investment $C^{\mathrm{unit}}$ over an $N$-year life at discount rate $r$; dividing by 365 gives the daily capital charge in $\mathrm{NV}(n)$. Higher $r$ or shorter $N$ therefore increases the effective BESS capital cost.

\subsubsection{Technology Comparison and Optimal Sizing}
\Cref{fig:battery_sensitivity_main} plots NV as a function of \revise{installed BESS energy capacity} for both battery technologies under each cycling limit at discount rates $r{=}7\%$ and $r{=}10\%$, and project lifetime $N= 20$ years. At $\bar N_{\text{cyc}}{=}1.0$ and $r{=}7\%$, the 2-hour Megapack~2XL peaks at 19~units (74.1~MWh/36.1~MW) with corrected $\mathrm{VA}{=}\$6{,}181$/day and $\mathrm{NV}{=}\$285$/day, whereas the 4-hour Megapack~3 reaches corrected $\mathrm{VA}{=}\$7{,}756$/day and $\mathrm{NV}{=}\$2{,}584$/day at 20~units (100~MWh/25~MW). At $r{=}10\%$, the optimal 2XL fleet collapses to one unit (3.9~MWh/1.9~MW) with $\mathrm{NV}{=}\$45$/day, while Megapack~3 remains at the upper search boundary with $\mathrm{NV}{=}\$1{,}320$/day. This difference reflects Megapack~3's substantially lower energy-specific capital cost (\$200/kWh versus \$308/kWh for 2XL).

The corrected first-unit value at $\bar N_{\text{cyc}}{=}1.0$ is similar on an energy basis, about \$111/MWh-day for Megapack~2XL and \$109/MWh-day for Megapack~3. At the $\bar N_{\text{cyc}}{=}1.0$, $r{=}7\%$ optima, AS revenue accounts for about 44\% of the 2XL's corrected value added (\$36/MWh-day) and 42\% of Megapack~3's corrected value added (\$32/MWh-day); the balance comes from non-AS operational flexibility. Thus, the technology ranking is not determined by energy duration alone: Megapack~2XL's higher power-to-energy ratio (0.49~MW/MWh versus 0.25~MW/MWh) supports ancillary service participation, but Megapack~3's lower capital cost yields higher net value.

\begin{figure}[t]
\centering
\vspace{-2em}
\includegraphics[width=.6\columnwidth]{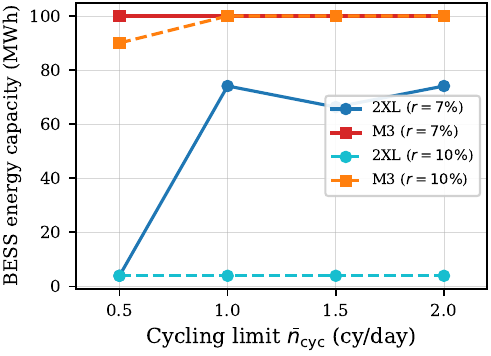}
\vspace{-1.2em}
\caption{Optimal installed BESS energy capacity versus cycling limit for discount rates $r{=}7\%$ and $r{=}10\%$. Each point is the tested fleet size that maximizes mean NV, computed from paired same-day robust objectives and averaged over $D=31$ days. Megapack~3 remains at or near the search boundary, while Megapack~2XL is more sensitive to cycling and discount-rate assumptions because its higher power density is paired with higher energy-specific capital cost.}
\vspace{-1.5em}
\label{fig:optimal_cycling}
\end{figure}

\subsubsection{Cycling-Limit Sensitivity}
\label{sec:cycling_sensitivity}
\Cref{fig:optimal_cycling} shows the optimal BESS capacity as a function of the cycling constraint. For Megapack~3, the most value is captured by {relaxing} $\bar N_{\text{cyc}}$ from 0.5 to 1.0~cy/day: at $r{=}7\%$, NV rises from about \$1.77k/day to \$2.58k/day, with little gain thereafter. Megapack~2XL is more sensitive: at $r{=}7\%$, optimal capacity increases from 3.9~MWh at 0.5~cy/day to 74.1~MWh at 1.0~cy/day. At $r{=}10\%$, however, one unit remains optimal across all cycling limits. Overall, about one cycle per day captures most of the value for Megapack~3, while 2XL economics depend more strongly on cycling and capital cost.

\subsection{Summary}
\label{sec:case_studies_summary}

The case studies show three complementary insights: i) the proposed co-optimization framework can deliver fully feasible day-ahead schedules with strong worst-case economic performance on realistic data. The commercial sizing results further show that power capability is valuable for flexibility services, but capital cost and cycling allowance can dominate the net-sizing decision. ii) A more flexible workload mix and BESS availability reduces the marginal impact of peak-load and ramping limits, but that effect is concentrated in the binding regime where net operating cost increases by greater than 40\%. In our case studies for stressed peak-load limits, increasing the fraction of schedulable workloads can reduce net operating cost by almost 20\%, while the daily value of BESS is doubled. In contrast, the value of relaxing ramp constraints, while also significant in the binding regime, remains largely unchanged across fixed/schedulable load compositions.
iii) Processor-level DVFS is an economically relevant control lever under tight load limits, reducing daily operating cost and improving schedulable-workload completion. Minimum market-participation commitments are inexpensive when pooled across products: an 8~MW aggregate floor raises mean net operating cost by only 0.4\% while retaining 100\% workload completion. By contrast, reserve-only commitments are infeasible on every day at 4~MW, whereas the aggregate floor remains feasible on all days through 10~MW.

Supplementary ablation studies in \cref{app:additional-ablations} assess LP-relaxation tightness, demand scaling, and out-of-sample robustness, finding a small relaxation gap and low load- and ramp-limit violation rates on unseen scenarios.

\section{Conclusion}
\label{sec:conclusion}
This paper examined how data centers can provide grid flexibility and services while maintaining the quality of service of their compute workloads. We developed a flexibility-centric robust day-ahead scheduling framework that endogenizes interconnection limits and ancillary service
market-participation constraints, so that committed flexibility is expressed in a form grid operators can directly consume and is guaranteed to be delivered under uncertainty. The formulation jointly coordinates workload scheduling, DVFS,
and co-located BESS dispatch in a tractable mixed-integer linear program. Using a 100~MW data center with CAISO and PJM data, we find that workload scheduling provides the largest flexibility volume, BESS mitigates short-duration peak and ramp constraints, and DVFS adds headroom under tight load limits. Binding interconnection constraints increase operating cost, particularly under tighter ramp limits, while the value of BESS and flexible workload composition rises when these constraints bind. Moderate mandatory ancillary service participation can be accommodated with limited impact on workload completion, although higher requirements can reduce service quality or become infeasible. Overall, the results show that coordinated {data center} resources can provide reliable grid-facing flexibility without materially compromising compute service, provided that interconnection constraints and workload requirements are explicitly accounted for.

Future work will extend the framework to include explicit models of cooling load and corrections to day-ahead schedules using model predictive control in real-time deployment. Further, we will integrate the proposed model and the resulting flexibility insights into data center planning problems.


\bibliographystyle{IEEEtran}
\Urlmuskip=0mu plus 1mu\relax
\bibliography{mainbib}

\newpage
\appendices
\crefalias{section}{appendix}

\onecolumn
\section{Explicit formulation}
\label{app:explicit-formulation}
{This appendix writes out \cref{eq:full-problem} in explicit form, with the named objective and constraint blocks replaced by their definitions from \Cref{sec:grid interface,sec:capacity-revenue,sec:energy-cost,sec:scheduling_frameworks,sec:workload-penalty,sec:bess_model,sec:cycling-cost}. The problem reads}
\begin{align*}
\min_{\substack{
P^{\mathrm{var}}_{j,t},\, z_{j,t,\ell},\, \widetilde{W}_j,\\
P^{\mathrm{B}}_{\alpha,t},\, P^{\mathrm{B}}_{\beta,t},\, z^{\alpha}_t,\, z^{\beta}_t,\\
P^{\mathrm{B},Res}_t,\, P^{\mathrm{B},\uparrow}_t,\, P^{\mathrm{B},\downarrow}_t,\\
P^{\mathrm{var}}_t,\, P^{\mathrm{DC}}_{t,s},\, D_{t,s},\\
E^{\mathrm{B}}_{s,t},\, E^{\mathrm{th}}_s,\, \widetilde{E}^{\mathrm{th}}_s
}}\;\;&
        \sum_{j\in\mathcal{J}}\sum_{t\in\mathcal{T}}\sum_{\ell\in\mathcal{L}}\!c^{\mathrm{DVFS}}_{j,\ell}\,z_{j,t,\ell} + \sum_{j\in\mathcal{J}}\!c^{\mathrm{unf}}_j\widetilde{W}_j + \sum_{j\in\mathcal{J}}\sum_{\substack{t\in\mathcal{T}\\ t>d_j}}\!\!\!c^{\mathrm{tardy}}_j(t-d_j)P^{\mathrm{var}}_{j,t}\Delta t {}-\sum_{t\in\mathcal{T}}\!\bigl[\lambda^{Res}_t P^{\mathrm{B},Res}_t + \lambda^{\uparrow}_t P^{\mathrm{B},\uparrow}_t + \lambda^{\downarrow}_t P^{\mathrm{B},\downarrow}_t\bigr]\Delta t\\[-8ex] 
        &{}+ \max_{s\in\mathcal{S}}\Bigl\{\sum_{t\in\mathcal{T}}\!\lambda^E_t\Delta t\bigl[D_{t,s} - \phi^{Res}_{t,s} P^{\mathrm{B},Res}_t + \phi^{\uparrow}_{t,s} P^{\mathrm{B},\uparrow}_t - \phi^{\downarrow}_{t,s} P^{\mathrm{B},\downarrow}_t\bigr] + c^{\mathrm{deg}}\widetilde{E}^{\text{th}}_s\Bigr\} \\\\
\st\;& D_{t,s} = P^{\mathrm{B}}_{\alpha,t} - P^{\mathrm{B}}_{\beta,t} + P^{\mathrm{DC}}_{t,s}, \\
& D_{t,s}\le\bar{D},\quad -\bar{\Delta}\le D_{t,s}-D_{t-1,s}\le\bar{\Delta}, \\
& P^{\mathrm{DC}}_{t,s} = P^{\mathrm{fixed}}_{t,s} + P^{\mathrm{var}}_t,\quad P^{\mathrm{DC}}_{t,s}\le\bar{P}^{\mathrm{DC}},\quad P^{\mathrm{var}}_t = \!\sum_{j\in\mathcal{J}}\!P^{\mathrm{var}}_{j,t}, \\[-1ex]
& 0\le P^{\mathrm{var}}_{j,t}\le \sum_{\ell\in\mathcal{L}}\overline{P}^{\mathrm{var}}_{j,\ell}\,z_{j,t,\ell},\quad P^{\mathrm{var}}_{j,t}=0\ \ \forall t<S_j, \\
& W_j-\widetilde{W}_j \le \!\sum_{t\in\mathcal{T}}\!P^{\mathrm{var}}_{j,t}\Delta t \le W_j,\quad \widetilde{W}_j\ge 0, \\[-1ex]
& \sum_{\ell\in\mathcal{L}}z_{j,t,\ell}\le 1,\quad z_{j,t,\ell}\in\{0,1\}, \\
& E^{\mathrm{B}}_{s,t} = E^{\mathrm{B}}_{s,t-1} + \bigl(P^{\mathrm{B}}_{\alpha,t}+\phi^{\uparrow}_{t,s} P^{\mathrm{B},\uparrow}_t\bigr)\eta^{\mathrm{B}}_{\alpha}\Delta t - \bigl(P^{\mathrm{B}}_{\beta,t}+\phi^{\downarrow}_{t,s} P^{\mathrm{B},\downarrow}_t + \phi^{Res}_{t,s} P^{\mathrm{B},Res}_t\bigr)\Delta t/\eta^{\mathrm{B}}_{\beta}, \\
& 0\le z^{\alpha}_t+z^{\beta}_t\le 1,\quad z^{\alpha}_t,z^{\beta}_t\in\{0,1\}, \\
& {P^{\mathrm{B}}_{\alpha,t}, P^{\mathrm{B},\uparrow}_t \ge 0,\;\;} P^{\mathrm{B}}_{\alpha,t}+P^{\mathrm{B},\uparrow}_t \le z^{\alpha}_t P^{\mathrm{B},\max}_{\alpha}, \\
& {P^{\mathrm{B}}_{\beta,t}, P^{\mathrm{B},Res}_t, P^{\mathrm{B},\downarrow}_t \ge 0,\;\;} P^{\mathrm{B}}_{\beta,t}+P^{\mathrm{B},Res}_t+P^{\mathrm{B},\downarrow}_t \le z^{\beta}_t P^{\mathrm{B},\max}_{\beta}, \\
& \underline{E}^{\mathrm{B}}\le E^{\mathrm{B}}_{s,t}\le \overline{E}^{\mathrm{B}}, \\
& E^{\text{th}}_s = \!\sum_{t\in\mathcal{T}}\!\Delta t\bigl(P^{\mathrm{B}}_{\alpha,t}+\phi^{\uparrow}_{t,s} P^{\mathrm{B},\uparrow}_t+P^{\mathrm{B}}_{\beta,t} +\phi^{Res}_{t,s} P^{\mathrm{B},Res}_t+\phi^{\downarrow}_{t,s} P^{\mathrm{B},\downarrow}_t\bigr), \\
& E^{\text{th}}_s \le \overline{E}^{\text{th}}+\widetilde{E}^{\text{th}}_s,\quad \widetilde{E}^{\text{th}}_s\ge 0, \\
& \forall t\in\mathcal{T},\ s\in\mathcal{S},\ j\in\mathcal{J},\ \ell\in\mathcal{L}.
\end{align*}
The full nomenclature used throughout the formulation is summarized in \crefrange{tab:nomenclature_compute}{tab:nomenclature_bess} below. Unless stated otherwise, power quantities are in MW, energy quantities are in MWh, and time is indexed in one-hour intervals.

\begin{table*}[!ht]
\caption{\revise{Nomenclature for compute load, workload-scheduling, and DVFS quantities.}}
\label{tab:nomenclature_compute}
\centering
\begingroup
\scriptsize
\renewcommand{\arraystretch}{1.15}
\begin{tabular}{p{0.18\textwidth}p{0.12\textwidth}p{0.64\textwidth}}
\hline
Symbol & Type & Meaning \\
\hline
$P^{\mathrm{fixed}}_{t,s}$ & Scenario data & Scenario-$s$ fixed job power load forecast (inelastic, DVFS-independent). \\
{$z_{j,t,\ell}$} & Decision variable & {Binary selector indicating whether job $j$ runs at DVFS level $\ell$ in period $t$. At most one $\ell$ is active per $(j,t)$; all-zero indicates job $j$ is idle in period $t$.} \\
$\bar P^{\mathrm{DC}}$ & Parameter & Internal server-side hardware power limit for the computing fleet. \\
$\mathrm{Flexible~Headroom}_{t,s}$ & Auxiliary variable & Residual internal capacity for schedulable jobs after reserving the fixed-load term. \\
$S_j$, $d_j$ & Parameter & Release time and deadline of schedulable job $j$. \\
$W_j$ & Parameter & Total work requirement of schedulable job $j$. \\
{$\overline{P}^{\mathrm{var}}_{j,\ell}$} & Parameter & {Maximum execution rate of schedulable job $j$ at DVFS level $\ell$.} \\
$P^{\mathrm{var}}_{j,t}$ & Decision variable & DVFS-weighted execution rate of job $j$ at time $t$ {(effective service after applying the selected DVFS level)}. \\
$\widetilde{W}_j$ & Decision variable & Unfinished work of schedulable job $j$ over the planning horizon. \\
{$C^{\mathrm{job}}$} & Auxiliary variable & {Aggregate workload-quality penalty ($C^{\mathrm{DVFS}}+C^{\mathrm{unf}}+C^{\mathrm{tardy}}$).} \\
{$c^{\mathrm{DVFS}}_{j,\ell}$} & Parameter & {Per-job, per-level DVFS-deviation penalty for selecting DVFS level $\ell$ on job $j$; $c^{\mathrm{DVFS}}_{j,0}=0$.} \\
{$c^{\mathrm{unf}}_j$} & Parameter & {Value of lost compute load for job $j$ (per unit of unfinished work).} \\
{$c^{\mathrm{tardy}}_j$} & Parameter & {Per-job, per-unit penalty for post-deadline execution, scaled by lateness.} \\
\hline
\end{tabular}
\endgroup
\end{table*}

\begin{table*}[!t]
\caption{\revise{Nomenclature for indices, grid interface quantities, and market quantities.}}
\label{tab:nomenclature_system}
\centering
\begingroup
\scriptsize
\renewcommand{\arraystretch}{1.15}
\begin{tabular}{p{0.18\textwidth}p{0.12\textwidth}p{0.64\textwidth}}
\hline
Symbol & Type & Meaning \\
\hline
$\mathcal{T}=\{1,\ldots,T\}$, $t$ & Index & Hourly scheduling horizon and time-period index. \\
$\Delta t$ & Parameter & Length of one time interval; equal to one hour in the case studies. \\
{$\mathcal{S}$, $s$} & Index & {Index set for the finite uncertainty trajectories and the associated scenario index.} \\
{$\xi_s$, $\Xi$} & Scenario data/set & {Joint trajectory of reserve deployment, upward/downward FRP deployment, and fixed-load forecast error; $\Xi=\{\xi_s:s\in\mathcal S\}$.} \\
{$\widehat P^{\mathrm{fixed}}_t$, $\varepsilon_{t,s}$} & Parameter/scenario data & {Fixed-load point forecast and additive scenario error, respectively.} \\
$\mathcal{J}$, $j$ & Index & Set of schedulable jobs and job index. \\
{$\mathcal{L}=\{0,\ldots,L\}$, $\ell$} & Index & {Discrete DVFS operating-level set and level index; $\ell=0$ is the nominal level.} \\
$P^{\mathrm{DC}}_{t,s}$ & Auxiliary variable & Total scenario-dependent compute power of the data center before BESS interaction. \\
$D_{t,s}$ & Auxiliary variable & Net grid withdrawal at the point of interconnection after BESS charge/discharge. Positive values denote grid consumption. \\
$\bar D$ & Parameter & Interconnection peak-load limit. \\
$\bar\Delta$ & Parameter & Interconnection ramp rate limit imposed on $D_{t,s}-D_{t-1,s}$. \\
$P^{\mathrm{B}}_{\alpha,t}$, $P^{\mathrm{B}}_{\beta,t}$ & Decision variable & Base BESS charging and discharging power schedules. \\
$P^{\mathrm{B},Res}_{t}$ & Decision variable & BESS reserve capacity offer. \\
$P^{\mathrm{B},\uparrow}_{t}$, $P^{\mathrm{B},\downarrow}_{t}$ & Decision variable & BESS upward and downward flexible ramping capacity offers. \\
$\lambda^E_t$ & Parameter & Energy price. \\
$\lambda^{Res}_t$ & Parameter & Reserve capacity price. \\
$\lambda^{\uparrow}_t$, $\lambda^{\downarrow}_t$ & Parameter & Upward and downward flexible ramping capacity prices. \\
$\phi^{Res}_{t,s}$ & Scenario data & Scenario-dependent reserve deployment factor. \\
$\phi^{\uparrow}_{t,s}$, $\phi^{\downarrow}_{t,s}$ & Scenario data & Scenario-dependent upward and downward flexible ramping deployment factors. \\
$R^{\mathrm{cap}}$ & Auxiliary variable & Total day-ahead capacity revenue ($R^{Res} + R^{\uparrow\downarrow}$). \\
$R^{Res}$, $R^{\uparrow\downarrow}$ & Auxiliary variable & Day-ahead reserve and flexible ramping capacity revenues. \\
$C^E_s$ & Auxiliary variable & Scenario-dependent realized energy settlement cost. \\
\hline
\end{tabular}
\endgroup
\end{table*}

\begin{table*}[!ht]
\caption{\revise{Nomenclature for BESS operation, degradation, and value calculations.}}
\label{tab:nomenclature_bess}
\centering
\begingroup
\scriptsize
\renewcommand{\arraystretch}{1.15}
\begin{tabular}{p{0.18\textwidth}p{0.12\textwidth}p{0.64\textwidth}}
\hline
Symbol & Type & Meaning \\
\hline
$E^{\mathrm{B}}_{s,t}$ & Decision variable & Scenario-dependent stored energy of the BESS. \\
$z^{\alpha}_t$, $z^{\beta}_t$ & Decision variable & BESS charge-mode and discharge-mode indicators. \\
$P^{\mathrm{B},\max}_{\alpha}$, $P^{\mathrm{B},\max}_{\beta}$ & Parameter & Maximum BESS charging and discharging power. \\
$\eta^{\mathrm{B}}_{\alpha}$, $\eta^{\mathrm{B}}_{\beta}$ & Parameter & BESS charging and discharging efficiencies. \\
$\underline{E}^{\mathrm{B}}$, $\overline{E}^{\mathrm{B}}$ & Parameter & Lower and upper admissible BESS stored-energy limits. \\
$E^{\mathrm{th}}_s$ & Auxiliary variable & Scenario-dependent total BESS energy throughput. \\
$\overline{E}^{\mathrm{th}}$ & Parameter & Nominal warranty throughput budget (twice the BESS energy capacity times the preferred daily cycle limit). \\
$\widetilde{E}^{\text{th}}_s$ & Decision variable & Scenario-dependent excess BESS throughput beyond the warranty budget. \\
$c^{\mathrm{deg}}$ & Parameter & Marginal BESS degradation cost per unit of excess throughput. \\
$C^{BESS}_s$ & Auxiliary variable & Scenario-dependent BESS cycling cost. \\
$n$ & Index & Number of identical commercial BESS units in the sizing study. \\
$C^{\mathrm{unit}}$ & Parameter & Capital cost per commercial BESS unit. \\
$r$, $N$ & Parameter & Discount rate and project lifetime used in the capital recovery factor. \\
$\mathrm{CRF}(r,N)$ & Parameter & Capital recovery factor converting investment cost into annualized cost. \\
$Z_d^\star(n)$ & Metric & Optimized daily operating cost on day $d$ with $n$ BESS units. \\
$D_{\mathrm{days}}$ & Parameter & Number of daily instances used when averaging value-added quantities. \\
$R^{\mathrm{AS}}$ & Auxiliary variable & Aggregate ancillary service revenue, $R^{Res}+R^{\uparrow\downarrow}$. \\
$V^{\mathrm{dep}}(n)$ & Metric & SOC-depletion correction applied to storage value-added calculations. \\
$\bar e_T$, $e_0$ & Metric & Mean terminal SOC fraction and initial SOC fraction used in SOC-depletion correction. \\
$\bar\lambda_e$ & Metric & Average energy price used in SOC-depletion correction. \\
$\mathrm{VA}(n)$, $\mathrm{VA}^{\mathrm{NA}}$ & Metric & Corrected total value added and corrected non-ancillary service value component. \\
$\mathrm{NV}(n)$ & Metric & Net value after subtracting annualized BESS capital cost. \\
$\lambda^{\mathrm{load}}_t$, $\lambda^{\mathrm{ramp\text{-}up}}_t$, $\lambda^{\mathrm{ramp\text{-}down}}_t$ & Metric & Shadow prices of load, upward-ramp, and downward-ramp constraints after fixing MILP integer variables. \\
\hline
\end{tabular}
\endgroup
\end{table*}

\makeatletter
\setlength{\@fptop}{0pt}
\makeatother
\twocolumn

\section{Schedulable Job Portfolio}
\label{app:job-portfolio}

\revise{The case studies use the schedulable job portfolio reported in \Cref{tab:job_portfolio}. The entries should be interpreted as aggregate schedulable workload classes, not typical detailed individual cluster jobs. This aggregation is appropriate for the present day-ahead grid co-optimization problem because the grid-facing model focuses on hourly flexible energy, release-deadline windows, maximum executable rates, and job-specific service-quality penalties rather than task-level identities. It is also consistent with large production traces, where clusters run very large numbers of heterogeneous jobs and applications that are commonly grouped by latency sensitivity, service/batch type, or delay tolerance~\cite{verma2015borg,cortez2017resource,cheng2018alibaba}. Each aggregate job is characterized by its release time $S_j$, deadline $d_j$, total work requirement $W_j$, maximum executable rate $\overline{P}^{\mathrm{var}}_j$, unfinished-work penalty $c^{\mathrm{unf}}_j$, and tardy-execution penalty $c^{\mathrm{tardy}}_j$. The planning horizon contains 24 one-hour periods, so $S_j$ denotes the first hour in which the aggregate job may be scheduled and $d_j$ denotes the last hour by which it should be completed. Work quantities are reported in MWh-equivalent compute-service units over the one-hour discretization, and rate limits are reported in MW-equivalent compute-service units.}

\begin{table*}[t]
\caption{\revise{Aggregate schedulable job portfolio used in the representative 100~MW case study. Work \(W_j\) is in MWh-equivalent compute-service units, maximum rate \(\overline{P}^{\mathrm{var}}_j\) is in MW-equivalent compute-service units, and \(c^{\mathrm{unf}}_j\) and \(c^{\mathrm{tardy}}_j\) are in \$/MWh and \$/(MWh$\cdot$h), respectively.}}
\label{tab:job_portfolio}
\centering
\begin{tabular*}{0.96\textwidth}{@{\extracolsep{\fill}}lcccccc}
\hline
Job class & Release $S_j$ & Deadline $d_j$ & Work $W_j$ & Max rate $\overline{P}^{\mathrm{var}}_j$ & $c^{\mathrm{unf}}_j$ & $c^{\mathrm{tardy}}_j$ \\
\hline
\texttt{batch\_analytics\_1} & 1  & 8  & 30.0 & 6.0  & 1600 & 80 \\
\texttt{hpc\_simulation}     & 3  & 14 & 60.0 & 10.0 & 4000 & 200 \\
\texttt{ml\_training\_large} & 6  & 20 & 70.0 & 9.0  & 5000 & 250 \\
\texttt{data\_backup}        & 1  & 24 & 25.0 & 5.0  & 1000 & 50 \\
\texttt{batch\_analytics\_2} & 12 & 20 & 30.0 & 7.0  & 2000 & 100 \\
\texttt{ml\_training\_small} & 10 & 18 & 20.0 & 5.0  & 3000 & 150 \\
\texttt{hpc\_evening}        & 16 & 24 & 40.0 & 10.0 & 3600 & 180 \\
\texttt{data\_processing}    & 1  & 12 & 35.0 & 4.0  & 4400 & 220 \\
\texttt{preemptable}         & 1  & 24 & 30.0 & 5.0  & 200  & 10 \\
\hline
\end{tabular*}
\end{table*}

\section{Additional Ablation Results}
\label{app:additional-ablations}

To complement the main-text case studies, this appendix reports the LP-relaxation study and additional multiday analyses of schedulable-workload headroom and out-of-sample robustness. Unless otherwise noted, all values are means and standard deviations over 31 daily instances.
\subsection{LP Relaxation and Computational Tightness}
\label{app:lp-relaxation}

Let $Z_{\mathrm{MILP}}$ denote the optimal value of the cost-minimization problem. Its direct LP relaxation replaces $z_{j,t,\ell}\in\{0,1\}$ in \cref{eq:dvfs-select} and $z_t^{\alpha},z_t^{\beta}\in\{0,1\}$ in \cref{eq:B_binary} by continuous variables on $[0,1]$, leaving all other constraints unchanged. The resulting value $Z_{\mathrm{LP}}\le Z_{\mathrm{MILP}}$ is a lower bound. A fractional DVFS selector is operationally admissible only under the heterogeneous-server interpretation in \cref{rem:dvfs-integrality}; moreover, fractional BESS modes may allocate capacity simultaneously to the charge and discharge sides. The relaxation is therefore used for screening and bounding, not as a directly implementable dispatch.

Using Gurobi~13, we evaluate the direct relaxation on the implemented seven-level case-study model, in which one common DVFS level is selected across jobs in each period, over 31 daily instances, each containing 50 uncertainty scenarios and a 12~MW/36~MWh BESS. The LP reduced mean solve time from 5.07 to 0.30~s (mean paired speedup: 18.4$\times$) and produced a mean net-operating-cost lower bound of $\$172.8\mathrm{k}\pm\$6.2\mathrm{k}$/day, compared with $\$172.9\mathrm{k}\pm\$6.2\mathrm{k}$/day for the MILP. The paired gap was $\$0.10\mathrm{k}\pm\$0.04\mathrm{k}$/day, or 0.060\% of the MILP objective magnitude; relaxing only the DVFS selectors and only the BESS modes accounted for $\$0.02\mathrm{k}\pm\$0.02\mathrm{k}$/day and $\$0.08\mathrm{k}\pm\$0.04\mathrm{k}$/day, respectively. Despite this small value gap, the full relaxation activated both BESS mode-capacity constraints in $7.0\pm2.1$ periods/day through simultaneous ancillary service offers. No simultaneous scheduled energy charging and discharging occurred in the base cases.

\subsection{Fixed-Load and Schedulable-Demand Scaling}
\label{sec:ablation_headroom}
{We define the residual internal capacity available to schedulable jobs after reserving the conservative fixed load as}
\begin{equation}\label{eq:headroom}
\mathrm{Flexible~Headroom}_{t,s} := \bar{P}^{\mathrm{DC}} - P^{\mathrm{fixed}}_{t,s}.
\end{equation}
{This is an auxiliary interpretive quantity, not an additional constraint; at the nominal DVFS level ($z_{j,t,0}=1$ for all jobs and periods), it is the \emph{base flexible headroom}.}

\begin{figure}[t]
\centering
\includegraphics[width=\columnwidth]{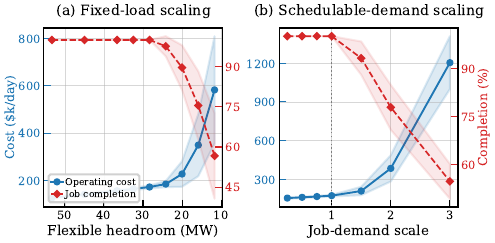}
\caption{Multiday fixed-load and schedulable-demand ablation. In panel~(a), scaling the fixed-load trace changes the residual headroom below the load cap; in panel~(b), scaling the schedulable workload leaves the fixed-load trace unchanged. \revise{In both panels, the blue curve denotes the mean daily operating cost and the red dashed curve denotes the mean job-completion rate, with shaded bands indicating \(\pm 1\) standard deviation over 31 daily instances.}}
\label{fig:ablation_headroom}
\end{figure}

\revise{\Cref{fig:ablation_headroom} contains two complementary sensitivity sweeps, and each panel reports both the daily operating cost (left axis) and the job-completion rate (right axis).} In \Cref{fig:ablation_headroom}(a), the fixed-load trace is scaled from 0.70 to 1.20, which changes the flexible headroom below the load limit (see \cref{eq:headroom}) from 53.3~MW to 11.7~MW. \revise{When the headroom is ample (roughly 28--53~MW), the completion rate remains at 100\% and the mean operating cost rises smoothly from \$127.3k\(\pm\$4.0\)k to \$173.1k\(\pm\$6.2\)k as fixed-load energy cost increases. Once the headroom falls below about 24~MW, however, performance deteriorates rapidly: at 20.0~MW headroom the mean operating cost rises to \$227.8k\(\pm\$53.7\)k with \(89.7\%\pm8.4\%\) completion, and at 11.7~MW headroom it rises to \$582.3k\(\pm\$226.5\)k with \(56.7\%\pm16.0\%\) completion.} 

\Cref{fig:ablation_headroom}(b) instead holds the fixed-load trace at the base case and scales the schedulable workload from 0.25\(\times\) to 3.0\(\times\), with per-job rate limits {$\overline{P}^{\mathrm{var}}_{j}$} increased proportionally up to 2.0\(\times\) as in the implementation. Here the average flexible headroom remains fixed at 28.3~MW over the scheduling horizon, so the degradation isolates the effect of growing schedulable demand. \revise{The system remains fully feasible through the base case, with the mean operating cost changing from \$154.2k\(\pm\$5.4\)k at 0.25\(\times\) demand to \$173.1k\(\pm\$6.2\)k at 1.0\(\times\). Beyond that point, the fixed headroom becomes insufficient for the enlarged job set: at 1.5\(\times\) demand the mean operating cost rises to \$209.1k\(\pm\$32.6\)k with \(93.2\%\pm5.2\%\) completion, at 2.0\(\times\) demand it becomes \$384.7k\(\pm\$101.8\)k with \(77.9\%\pm7.0\%\) completion, and at 3.0\(\times\) demand it rises to \$1.20M\(\pm\$205.5\)k with \(54.8\%\pm5.4\%\) completion.} Taken together, the two panels show that both fixed-load growth and schedulable-demand growth can push the system past a sharp feasibility threshold; the common driver is insufficient residual capacity available to serve schedulable jobs. This is similar to results in \Cref{fig:ic_sensitivity} where the load and ramp limits were progressively tightened.

\subsection{Out-of-Sample Robustness}

\begin{figure}[t]
\centering
\includegraphics[width=\columnwidth]{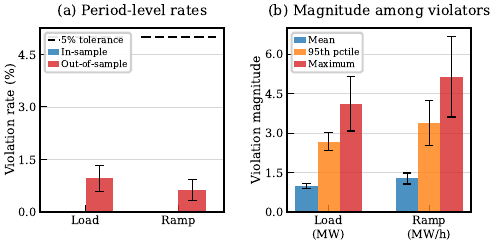}
\caption{Multiday out-of-sample robustness. The first-stage schedule from each 50-scenario training problem is evaluated on 500 fresh scenarios per day. Panel (a) reports period-level load-limit and ramp-limit violation rates, which remain well below the imposed 5\% tolerance. Panel (b) reports the magnitude of out-of-sample violations conditional on violation occurrence. Error bars indicate mean \(\pm\) standard deviation over 31 daily instances.}
\label{fig:ablation_oos}
\end{figure}

This analysis evaluates the first-stage solution from each 50-scenario trained solution on 500 new day-long scenarios of ancillary service (reserve and ramping) deployments, drawn from the same uncertainty model. As summarized in \Cref{fig:ablation_oos}, the in-sample violation rates are zero on average for both the load and ramp limits. Out of sample, the corresponding period/time-instance level violation rates remain low at \(0.96\%\pm0.38\%\) for load and \(0.63\%\pm0.30\%\) for ramp, both well below the 5\% tolerance used in the optimization. The mean maximum out-of-sample exceedances are 4.12~MW for load and 5.13~MW/h for ramp. Because a day-long scenario contains 24 hourly tests, the probability that a scenario contains at least one violation is naturally larger than the period-level rate; in our tests, these scenario-level incidences are \(20.4\%\) for load and \(13.7\%\) for ramp. These results indicate that the scenario-based robust formulation generalizes well for the interconnection-feasibility metrics that are enforced in the optimization model.

\end{document}

%% file: preamble.tex
\usepackage{enumitem}

\newcommand{\st}{\mathop{\text{\normalfont s.t.}}}

\usepackage{braket,amsfonts}

\usepackage{array}

\usepackage[caption=false]{subfig}
\usepackage{pgfplots}
\pgfplotsset{compat=1.17}

\usepackage{amsthm,amsmath}

\usepackage[colorlinks = true,
            linkcolor = blue,
            urlcolor  = blue,
            citecolor = blue,
            anchorcolor = blue]{hyperref}
\usepackage[nameinlink]{cleveref}
\crefname{assumption}{assumption}{assumptions}
\crefname{equation}{}{}
\Crefname{equation}{Equation}{Equations}
\crefname{appendix}{Appendix}{Appendices}
\Crefname{appendix}{Appendix}{Appendices}
\crefname{table}{Table}{Tables}
\crefname{figure}{Fig.}{Figs.}
\crefname{remark}{Remark}{Remarks}
\crefrangeformat{equation}{#3(#1)#4-#5(#2)#6}
\Crefrangeformat{equation}{#3(#1)#4-#5(#2)#6}

\newtheorem{remark}{Remark}

\usepackage{graphicx,epstopdf}

\usepackage{amsopn}

\usepackage{xspace}
\usepackage{bold-extra}
\usepackage[most]{tcolorbox}

\colorlet{texcscolor}{blue!50!black}
\colorlet{texemcolor}{red!70!black}
\colorlet{texpreamble}{red!70!black}
\colorlet{codebackground}{black!25!white!25}

\lstdefinestyle{siamlatex}{%
  style=tcblatex,
  texcsstyle=*\color{texcscolor},
  texcsstyle=[2]\color{texemcolor},
  keywordstyle=[2]\color{texemcolor},
  moretexcs={cref,Cref,maketitle,mathcal,text,headers,email,url},
}

\tcbset{%
  colframe=black!75!white!75,
  coltitle=white,
  colback=codebackground, 
  colbacklower=white, 
  fonttitle=\bfseries,
  arc=0pt,outer arc=0pt,
  top=1pt,bottom=1pt,left=1mm,right=1mm,middle=1mm,boxsep=1mm,
  leftrule=0.3mm,rightrule=0.3mm,toprule=0.3mm,bottomrule=0.3mm,
  listing options={style=siamlatex}
}

\newtcblisting[use counter=example]{example}[2][]{%
  title={Example~\thetcbcounter: #2},#1}

\newtcbinputlisting[use counter=example]{\examplefile}[3][]{%
  title={Example~\thetcbcounter: #2},listing file={#3},#1}

\DeclareTotalTCBox{\code}{ v O{} }
{ 
  fontupper=\ttfamily\color{black},
  nobeforeafter,
  tcbox raise base,
  colback=codebackground,colframe=white,
  top=0pt,bottom=0pt,left=0mm,right=0mm,
  leftrule=0pt,rightrule=0pt,toprule=0mm,bottomrule=0mm,
  boxsep=0.5mm,
  #2}{#1}

\allowdisplaybreaks
\usepackage{mathtools}
\usepackage{parskip}


%% file: mainbib.bib
@article{lauinger2025value,
  title={The value of storage in electricity distribution: The role of markets},
  author={Lauinger, Dirk and Deka, Deepjyoti and Shin, Sungho},
  journal={arXiv preprint arXiv:2510.12435},
  year={2025}
}

@Misc{ercot2025plan,
  author = {ERCOT},
  title = {2025 {ERCOT} System Planning Long-Term Hourly Peak Demand and Energy Forecast},
  month = apr,
  year = 2025,
  address = {Texas, USA}  
}

@Misc{nerc2026gaps,
  author = {{North American Electric Reliability Corporation}},
  title = {Assessment of Gaps in Existing Practices, Requirements, and Reliability Standards for Emerging Large Loads},
  month = mar,
  year = 2026,
}

@article{zhang2020flexibility,
  title={Flexibility from networks of data centers: A market clearing formulation with virtual links},
  author={Zhang, Weiqi and Roald, Line A and Chien, Andrew A and Birge, John R and Zavala, Victor M},
  journal={Electric Power Systems Research},
  volume={189},
  pages={106723},
  year={2020},
  publisher={Elsevier}
}

@article{hasegawa2023demandresponse,
  author = {Hasegawa, Raiden},
  title = {Using demand response to reduce data center power consumption},
  journal = {Google Cloud Blog},
  year = {2023},
  month = {October},
  day = {3},
  url = {https://cloud.google.com/blog/products/infrastructure/using-demand-response-to-reduce-data-center-power-consumption}
}

@ARTICLE{zhao2016flex,
  author={Zhao, Jinye and Zheng, Tongxin and Litvinov, Eugene},
  journal={IEEE Transactions on Power Systems}, 
  title={A Unified Framework for Defining and Measuring Flexibility in Power System}, 
  year={2016},
  volume={31},
  number={1},
  pages={339-347},
  doi={10.1109/TPWRS.2015.2390038}}

@ARTICLE{Riaz2022flex,
  author={Riaz, Shariq and Mancarella, Pierluigi},
  journal={IEEE Transactions on Power Systems}, 
  title={Modelling and Characterisation of Flexibility From Distributed Energy Resources}, 
  year={2022},
  volume={37},
  number={1},
  pages={38-50},
  doi={10.1109/TPWRS.2021.3096971}}

@techreport{norris2025rethinking,
    author = {Norris, Tyler and Profeta, Timothy and Patino-Echeverri, Dalia and Cowie-Haskell, Adam},
    title = {Rethinking load growth: assessing the potential for integration of large flexible loads in US power systems},
    institution = {Nicholas Institute for Energy, Environment \& Sustainability},
    year={2025},
}

@article{ghaljehei2021day,
  title={Day-ahead operational scheduling with enhanced flexible ramping product: Design and analysis},
  author={Ghaljehei, Mohammad and Khorsand, Mojdeh},
  journal={IEEE transactions on power systems},
  volume={37},
  number={3},
  pages={1842--1856},
  year={2021},
  publisher={IEEE}
}

@article{wang2018adjustable,
  title={An adjustable chance-constrained approach for flexible ramping capacity allocation},
  author={Wang, Zhiwen and Shen, Chen and Liu, Feng and Wang, Jianhui and Wu, Xiangyu},
  journal={IEEE Transactions on Sustainable Energy},
  volume={9},
  number={4},
  pages={1798--1811},
  year={2018},
  publisher={IEEE}
}

@techreport{cole2025cost,
  title={Cost projections for utility-scale battery storage: 2025 update},
  author={Cole, Wesley and Ramasamy, Vignesh and Turan, Merve},
  year={2025},
  institution={National Renewable Energy Lab.(NREL), Golden, CO (United States)}
}

@article{fan2007power,
  title={Power provisioning for a warehouse-sized computer},
  author={Fan, Xiaobo and Weber, Wolf-Dietrich and Barroso, Luiz Andre},
  journal={ACM SIGARCH computer architecture news},
  volume={35},
  number={2},
  pages={13--23},
  year={2007},
  publisher={ACM New York, NY, USA}
}

@inproceedings{guenter2011managing,
  title={Managing cost, performance, and reliability tradeoffs for energy-aware server provisioning},
  author={Guenter, Brian and Jain, Navendu and Williams, Charles},
  booktitle={2011 Proceedings IEEE INFOCOM},
  pages={1332--1340},
  year={2011},
  organization={IEEE}
}

@techreport{balducci2019nantucket,
  title={Nantucket island energy storage system assessment},
  author={Balducci, Patrick J and Alam, Jan E and McDermott, Thomas E and Fotedar, Vanshika and Ma, Xu and Wu, Di and Bhatti, Bilal Ahmad and Mongird, Kendall and Bhattarai, Bishnu P and Crawford, Aladsair J and others},
  year={2019},
  institution={Pacific Northwest National Laboratory, Richland, WA (U.S.)}
}

@inproceedings{piga2024expanding,
  title={Expanding datacenter capacity with dvfs boosting: A safe and scalable deployment experience},
  author={Piga, Leonardo and Narayanan, Iyswarya and Sundarrajan, Aditya and Skach, Matt and Deng, Qingyuan and Maity, Biswadip and Chakkaravarthy, Manoj and Huang, Alison and Dhanotia, Abhishek and Malani, Parth},
  booktitle={Proceedings of the 29th ACM International Conference on Architectural Support for Programming Languages and Operating Systems, Volume 1},
  pages={150--165},
  year={2024}
}

@inproceedings{patel2024characterizing,
  title={Characterizing power management opportunities for {LLM}s in the cloud},
  author={Patel, Pratyush and Choukse, Esha and Zhang, Chaojie and Goiri, {\'I}{\~n}igo and Warrier, Brijesh and Mahalingam, Nithish and Bianchini, Ricardo},
  booktitle={Proc. of 29th ACM Int. Conf. on Arch. Support for Programming Languages and Operating Systems, Volume 3},
  year={2024},
  doi={10.1145/3620666.3651329}
}

@inproceedings{stojkovic2025tapas,
  title={{TAPAS}: Thermal- and power-aware scheduling for {LLM} inference in cloud platforms},
  author={Stojkovic, Jovan and Zhang, Chaojie and Goiri, {\'I}{\~n}igo and Choukse, Esha and Qiu, Haoran and Fonseca, Rodrigo and Torrellas, Josep and Bianchini, Ricardo},
  booktitle={Proceedings of the 30th ACM International Conference on Architectural Support for Programming Languages and Operating Systems, Volume 2},
  year={2025},
  doi={10.1145/3676641.3716025}
}

@article{hall2025carbon,
  title={Carbon-aware computing for data centers with probabilistic performance guarantees},
  author={Hall, Sophie and Micheli, Francesco and Belgioioso, Giuseppe and Radovanovi{\'c}, Ana and D{\"o}rfler, Florian},
  journal={IEEE Transactions on Power Systems},
  year={2025},
  publisher={IEEE}
}

@article{li2015flexible,
  title={Flexible operation of batteries in power system scheduling with renewable energy},
  author={Li, Nan and Uckun, Canan and Constantinescu, Emil M and Birge, John R and Hedman, Kory W and Botterud, Audun},
  journal={IEEE Transactions on Sustainable Energy},
  volume={7},
  number={2},
  pages={685--696},
  year={2015},
  publisher={IEEE}
}

@article{dvorkin2024agent,
  title={Agent coordination via contextual regression (agentconcur) for data center flexibility},
  author={Dvorkin, Vladimir},
  journal={IEEE Transactions on Power Systems},
  volume={40},
  number={2},
  pages={1832--1842},
  year={2024},
  publisher={IEEE}
}

@article{radovanovic2022carbon,
  title={Carbon-aware computing for datacenters},
  author={Radovanovi{\'c}, Ana and Koningstein, Ross and Schneider, Ian and Chen, Bokan and Duarte, Alexandre and Roy, Binz and Xiao, Diyue and Haridasan, Maya and Hung, Patrick and Care, Nick and others},
  journal={IEEE Transactions on Power Systems},
  volume={38},
  number={2},
  pages={1270--1280},
  year={2022},
  publisher={IEEE}
}

@inproceedings{lin2023adapting,
  title={Adapting datacenter capacity for greener datacenters and grid},
  author={Lin, Liuzixuan and Chien, Andrew A},
  booktitle={Proceedings of the 14th ACM International Conference on Future Energy Systems},
  pages={200--213},
  year={2023}
}

@inproceedings{guruprasad2017coupling,
  title={Coupling a small battery with a datacenter for frequency regulation},
  author={Guruprasad, Ranjini and Murali, Prakash and Krishnaswamy, Dilip and Kalyanaraman, Shivkumar},
  booktitle={2017 IEEE Power \& Energy Society General Meeting},
  year={2017},
  organization={IEEE}
}

@inproceedings{wierman2014opportunities,
  title={Opportunities and challenges for data center demand response},
  author={Wierman, Adam and Liu, Zhenhua and Liu, Iris and Mohsenian-Rad, Hamed},
  booktitle={International Green Computing Conference},
  pages={1--10},
  year={2014},
  organization={IEEE}
}

@article{fu2020assessments,
  title={Assessments of data centers for provision of frequency regulation},
  author={Fu, Yangyang and Han, Xu and Baker, Kyri and Zuo, Wangda},
  journal={Applied Energy},
  volume={277},
  year={2020},
  publisher={Elsevier}
}

@techreport{brancucci2025flexible,
  author       = {Carlo Brancucci and Dylan Cutler and Jesse Jenkins},
  title        = {Flexible Data Centers: A Faster, More Affordable Path to Power},
  institution  = {Camus and encoord and Princeton ZERO Lab},
  month        = dec,
  year         = {2025},
  type         = {Technical Report}
}

@inproceedings{verma2015borg,
  author    = {Verma, Abhishek and Pedrosa, Luis and Korupolu, Madhukar and Oppenheimer, David and Tune, Eric and Wilkes, John},
  title     = {Large-scale Cluster Management at {Google} with {Borg}},
  booktitle = {Proceedings of the Tenth European Conference on Computer Systems},
  series    = {EuroSys '15},
  year      = {2015},
  pages     = {1--17},
  publisher = {ACM},
  doi       = {10.1145/2741948.2741964}
}

@misc{aws2026p5pricing,
  author       = {{Amazon Web Services}},
  title        = {Amazon {EC2} Capacity Blocks for {ML} Pricing},
  year         = {2026},
  howpublished = {Online},
  url          = {https://aws.amazon.com/ec2/capacityblocks/pricing/},
  note         = {Accessed: Jul. 30, 2026}
}

@techreport{nvidia2023dgxh100,
  author      = {{NVIDIA Corporation}},
  title       = {{NVIDIA DGX H100}: The Gold Standard for {AI}},
  institution = {NVIDIA Corporation},
  type        = {Datasheet},
  year        = {2023},
  url         = {https://www.nvidia.com/content/dam/en-zz/Solutions/Data-Center/nvidia-dgx-h100-datasheet.pdf}
}

@inproceedings{cortez2017resource,
  author    = {Cortez, Eli and Bonde, Anand and Muzio, Alexandre and Russinovich, Mark and Fontoura, Marcus and Bianchini, Ricardo},
  title     = {Resource Central: Understanding and Predicting Workloads for Improved Resource Management in Large Cloud Platforms},
  booktitle = {Proceedings of the 26th Symposium on Operating Systems Principles},
  series    = {SOSP '17},
  year      = {2017},
  pages     = {153--167},
  publisher = {ACM},
  doi       = {10.1145/3132747.3132772}
}

@inproceedings{cheng2018alibaba,
  author    = {Cheng, Yue and Anwar, Ali and Duan, Xuejing},
  title     = {Analyzing {Alibaba}'s Co-located Datacenter Workloads},
  booktitle = {Proceedings of the 2018 IEEE International Conference on Big Data},
  year      = {2018},
  pages     = {292--297},
  publisher = {IEEE},
  doi       = {10.1109/BigData.2018.8622518}
}

@inproceedings{googletrace2020,title	= {Data Center Power Oversubscription with a Medium Voltage Power Plane and Priority-Aware Capping},author	= {Varun Sakalkar and Vasileios Kontorinis and David Landhuis and Shaohong Li and Darren De Ronde and Thomas Blooming and Anand Ramesh and James Kennedy and Christopher Malone and Jimmy Clidaras and Parthasarathy Ranganathan},year	= {2020},booktitle	= {Proc. of the 25th Int. Conf. on Architectural Support for Programming Languages and Operating Systems},pages	= {497–511},address	= {New York, NY, USA}}

@misc{caiso2025ramp,
  author       = {{CAISO}},
  title        = {California {ISO} Open Access Same-time Information System ({OASIS})},
  howpublished = {\url{https://oasis.caiso.com/mrioasis/logon.do}},
  note         = {Database, accessed 2025},
}

@misc{pjm2025reserve,
  author       = {{PJM}},
  title        = {{PJM} Data Miner 2},
  howpublished = {\url{https://dataminer2.pjm.com/list}},
  note         = {Database, accessed 2025},
}

@article{engels2020integration,
  title={Optimal Combination of Frequency Control and Peak Shaving With Battery Storage Systems},
  author={Engels, Jonas and Claessens, Bert and Deconinck, Geert},
  journal={IEEE Transactions on Smart Grid},
  volume={11},
  number={4},
  pages={3270--3279},
  year={2020},
  doi={10.1109/TSG.2019.2963098},
  publisher={IEEE}
}

@techreport{stewart2024bess,
  author      = {Stewart, Emma Mary and Culler, Megan Jordan and Stolworthy, Remy Vanece},
  title       = {{Battery Energy Storage Systems Report}},
  institution = {{Idaho National Laboratory (INL)}},
  address     = {Idaho Falls, ID, United States},
  year        = {2024},
  month       = nov,
  doi         = {10.2172/3023305}
}
